%% file: paper.tex
\documentclass[10pt]{sig-alternate-05-2015} 
\usepackage{array,multirow}
\usepackage{nohyperref}
\usepackage{flushend}
\usepackage{graphicx} 
\usepackage[usenames, dvipsnames]{color} 
\usepackage{url} %
\usepackage{amsmath}
\usepackage{amssymb}
\usepackage{cite}
\usepackage{booktabs}
\usepackage{amssymb}
\usepackage{xspace}
\usepackage{times}
\usepackage{subfigure}
\usepackage{enumitem}

\newcommand\xref[1]{\S~\ref{#1}}

\newcommand{\neta}{Netalyzr\xspace}
\newcommand{\parax}[1]{\vspace{0.2em} \noindent \textbf{#1:}}

\DeclareFixedFont{\afacc}{OT1}{phv}{m}{n}{10}

\input{preamble.tex}

\begin{document}

\CopyrightYear{2016}
\setcopyright{acmlicensed}
\conferenceinfo{IMC 2016,}{November 14 - 16, 2016, Santa Monica, CA, USA}
\isbn{978-1-4503-4526-2/16/11}\acmPrice{\$15.00}
\doi{http://dx.doi.org/10.1145/2987443.2987474}

\clubpenalty=10000 
\widowpenalty = 10000

\title{A Multi-perspective Analysis of\\ Carrier-Grade NAT Deployment}

\numberofauthors{1}
\author{
\alignauthor
  Philipp Richter$^1$, Florian Wohlfart$^2$, Narseo Vallina-Rodriguez$^3$,\\ 
  \vspace{1mm}
  Mark Allman$^3$, Randy Bush$^5$, Anja Feldmann$^1$, Christian 
  Kreibich$^{3,6}$,\\ \vspace{1mm} 
  Nicholas Weaver$^{3}$, Vern Paxson$^{3,4}$ \\ \vspace{2mm}
  \affaddr{{$^1$}TU Berlin, {$^2$}TU M\"unchen, 
  {$^3$}ICSI, {$^4$}UC Berkeley, 
  {$^5$}Internet Initiative 
  Japan, {$^6$}Lastline}\\
}

\maketitle

\begin{abstract}

As ISPs face IPv4 address scarcity they increasingly turn to network
address translation (NAT) to accommodate the address needs of their
customers.  Recently, ISPs have moved beyond employing NATs only
directly at individual customers and instead begun deploying
\textit{Carrier-Grade NATs} (CGNs) to apply address translation to
many independent and disparate endpoints spanning physical locations,
a phenomenon that so far has received little in the way of empirical
assessment.
In this work we present a broad and systematic study of the
deployment and behavior of these middleboxes.
We develop a methodology to detect the existence of hosts behind CGNs
by extracting non-routable IP addresses from peer lists we obtain by
crawling the BitTorrent DHT. We complement this approach with
improvements to our Netalyzr troubleshooting service, enabling us to
determine a range of indicators of CGN presence as well as detailed insights 
into key properties of CGNs.
Combining the two data sources we illustrate the scope of CGN
deployment on today's Internet, and report on characteristics of commonly
deployed CGNs and their effect on end users.

\end{abstract}

\section{Introduction}
\label{sec:intro}

As originally designed, the Internet architecture calls for IP
addresses to uniquely identify devices.  This structure lays the
foundation for a peer-to-peer system that facilitates direct
communication between hosts.  However, this model runs into trouble
once addresses become scarce.  This situation first
manifested in home networks, where ISPs provided subscribers with only
a single IP address while the number of devices in
home networks ballooned.

Today, Network Address Translation (NAT)
\cite{rfc1631} is ubiquitous at the edge of home networks to meet
both the ISPs' desire to conserve IP addresses and the users'
requirement of connecting a multitude of devices.  IP
address scarcity has long moved beyond home networks and onto the
global stage~\cite{RABP14}.
Thus,
even though IPv6 is gaining momentum \cite{ipv6sigcomm14},
large ISPs are confronted by address shortages, as we
illustrate in \xref{sec:ops}%
, and hence
turn to a well-worn coping technique: NAT. 

Instead of aggregating small populations of tightly-knit users and
devices within one residence under a single IP address,
\textit{Carrier-Grade NATs} (CGNs) apply NAT to many independent and
disparate endpoints spanning physical locations.  On
one level we can view CGNs as representing a second instantiation of
a well-known technique for combating address shortages.  While
tempting, conflating CGNs with small edge-based NATs represents a
false equivalence, for two reasons: ($i$) by operating at large scales,
CGNs face issues not present in residential settings, which have
received more examination,
and ($ii$)
CGNs generally represent a second level of address
translation---i.e., CGNs operate \textit{in addition} to existing
edge-based NAT---and therefore compound some of the issues that address
translation raises.

While we know anecdotally that ISPs deploy CGN,
we are not aware of quantitative studies of the prevalence and
operation of CGNs in the wild.  In this work, we take a first step
toward developing an empirical understanding of these increasingly
crucial pieces of Internet infrastructure.  We make
four high-level contributions:

\parax{Operator Perspectives on CGNs}
We begin by presenting a survey of operators in \xref{sec:ops}.  We
distributed a questionnaire on pertinent mailing lists, seeking
to shed light on operators' motivations and experiences with CGN
operation in the wild.  We received illuminating input from
75~operators.
Our
survey reveals widespread adoption of CGN technology---with over
half of the responding operators having deployed CGNs or planning to in the near
future---despite the resulting operational difficulties.
\parax{Measurement Methodology}
One of the key characteristics of CGNs is their \textit{transparent} operation
from the perspective of endpoints.  While transparency has its
benefits (e.g., clients require no setup process to use a CGN), it
complicates detection and measurement of CGNs.
Multiple levels of address translation increase the difficulty
further as each step overwrites any evidence a previous NAT left in
the traffic.  Therefore, the sender of a packet cannot tell if or
how many times the source address will be translated on the path
towards a destination, and the recipient cannot know the original
source of the packet.

To address these difficulties we introduce two methods in
\xref{sec:cgndetection} for exploring CGNs.  
First, we observe that some nodes in the BitTorrent DHT mistake addresses 
internal to a CGN for external addresses and therefore propagate (``leak'') 
these to other nodes.  Therefore, we are able to derive a broad understanding of
the deployment of CGNs by probing the DHT.  Our second set of methods relies 
on extensions to our Netalyzr measurement platform~\cite{netalyzr:imc:short},
which allow us to study 
the presence and detailed properties of CGNs based on locally available 
addressing information, repeated connectivity tests,
as well as a new method that 
leverages the stateful nature of NATs and uses TTL-limited probes to force 
retention of state in some hops while allowing it to expire in
others.

\parax{Studying Global CGN Presence}
IPv4 address scarcity manifests differently for different networks
in different parts of the world \cite{RABP14}. Our CGN detection
methods give us a broad and unprecedented view into the global
deployment of CGNs, which we present in
\xref{sec:networkwideview}. Our vantage points cover more than 60\% of
the 
Internet's ``Eyeball ASes'' that connect end users to the Internet.
We find the CGN penetration rate to be 17--18\% of all Eyeball ASes.
Moreover, we find that CGN deployment is ubiquitous in cellular
networks with more than 90\% of all cellular ASes deploying CGNs.
We also find a direct relationship between regions with higher
perceived IPv4 address scarcity and CGN deployment.

\parax{Understanding CGN Behavior}
CGNs present a massive resource distribution problem, whereby scarce public 
IPv4 addresses are multiplexed using a relatively small set of internal IPv4 
addresses and a limited port space across thousands of end hosts. CGNs can be 
configured in a multitude of ways, with currently little known about CGN
configurations, 
dimensioning, and behavior in the wild. Hence, in \xref{sec:drilling} we
make our final
contribution: a deep dive into 
the properties of deployed CGNs. We analyze 
the internal address ranges used by CGNs, which reveals that some
ISPs even face scarcity of internally used (``private'') address space.  We 
also 
find CGN placement is diverse, ranging from
\mbox{1--12~hops}
from the user.  We find that the methods CGNs use to distribute
available public IP  
addresses and port numbers to their subscribers vary
dramatically.  We then assess how CGNs restrict user
connectivity and  
compare our insights about CGNs to the properties of commonly
deployed CPE (customer 
premises equipment) NATs.

Finally, we note that while our study provides an unprecedented view
into the use and properties of CGNs in the wild, we only partially
illuminate the CGN landscape.  Each of our measurement approaches has
limitations that somewhat restrict their scope.  For instance, since
mobile devices rarely use BitTorrent, our DHT crawl does not shed
significant light on the use of CGNs within mobile ISPs.  Our study
constitutes
an initial view into the deployment of CGNs with much future
work to be done to better understand the impact of these critical
components of the modern Internet.

\section{An Operator's Perspective}
\label{sec:ops}

To gain a better understanding of the real-world challenges that
IPv4 address scarcity poses and how ISPs are coping, in late 2015
we circulated a survey on a dozen of network operator mailing lists and 
eventually collected responses from 75 ISPs located all over the world.  These 
ISPs run the gamut in terms of size and type, including cellular and 
residential ones. While we do not claim the respondents form a statistically 
unbiased sample, we note that we received answers from operators in all regions 
of the world, spanning the whole spectrum of ISPs (cellular, residential) 
ranging from small rural ISPs in Africa up to Fortune 50 companies, connecting 
millions of subscribers to the Internet. Thus, we do believe that the 
approaches and concerns raised by these ISPs deserve our attention.  Next, we 
summarize the survey responses.

\parax{IPv4 Address Space Scarcity}
More than 40\% of the responding ISPs indicate that they directly
face IPv4 address scarcity issues.  Some ISPs report a
subscriber-to-IPv4 address ratio as high as 20:1.  However, others
point out that while their subscriber-to-address ratio is 1:1,
internal subnetting and fragmentation make address space management
cumbersome, especially when attempting to accommodate new customers.
Another 10\% of the respondents indicate that while they do not yet
face scarcity, they believe it is looming in the near future.  The
ISPs not facing IPv4 address scarcity are mainly ones that received
significant blocks of address space many years ago, as well as ISPs
in the African region.\footnote{Africa is the only region in which
  the IPv4 address pool is not yet depleted.}  Interestingly, three
ISPs also indicated that they face scarcity of \textit{internal} IPv4
address space.  These networks leverage CGN but also need internal
address space for their internal management.

\parax{IPv4 Address Space Markets}
Three of the responding ISPs report that they have bought IPv4
addresses, while another 15~ISPs indicate that they have considered
procuring additional addresses.  However, ISPs indicate
concern regarding buying address space, including price of
available address blocks (named by 60\%), fear of obtaining ``polluted'' 
address blocks with a bad reputation from previous use (44\%) and uncertainty 
regarding the ownership of blocks (42\%).

\begin{figure}
  \begin{center}
  \subfigure[Carrier-Grade NAT.]{
    \includegraphics[width=0.43\linewidth]{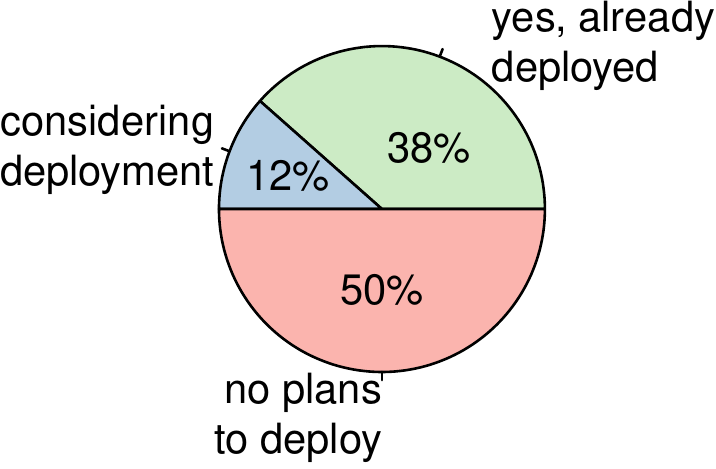}
    \label{fig:surveycgnpie}
  }
  \subfigure[IPv6.]{
    \includegraphics[width=0.47\linewidth]{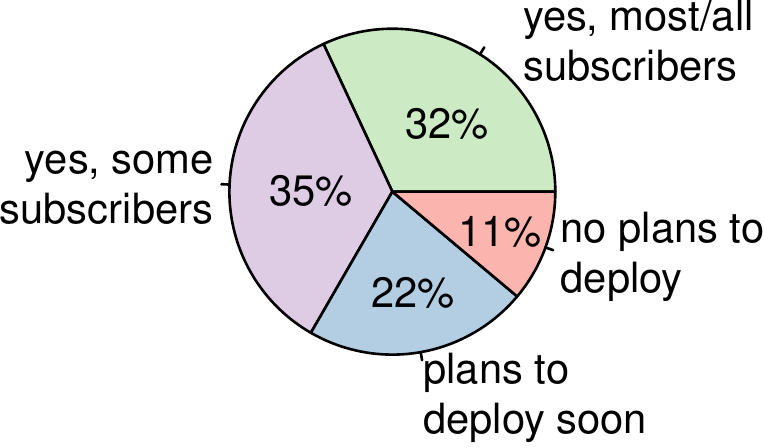}
    \label{fig:surveyv6pie}
  }
  \caption{ISP survey: Status of Carrier-Grade NAT deployment and IPv6 
  deployment.}
    \label{fig:ispsurvey_piechart}
  \end{center}
\vspace{-1em}
\end{figure}

\parax{CGN Deployment vs. IPv6 Deployment}
Figure \ref{fig:ispsurvey_piechart} shows the respondents' approach
to CGN and IPv6.  Almost 40\% of the ISPs indicate they deploy
IPv4 CGNs, with another 12\% considering CGN in the near-term.
Typically, ISPs note incremental CGN deployments, either targeting
new customers or shifting specific subsets of subscribers into CGN
deployment.  That is, \textit{most CGN deployments are partial}.
Next we find that 32\% of the ISPs indicate IPv6 deployment to most
or all of their subscribers, while another 35\% have partial IPv6
deployments for some subscribers.  The dominant transition mechanism
noted is dual stack.  Some ISPs also provide customers with an
internal (CGN) IPv4 address and a publicly reachable IPv6 address.
This arrangement will likely gain popularity in the near future as
IPv4 connectivity will remain necessary until full IPv6 deployment.

\parax{CGN Concerns}
Participating ISPs also had the option to inform us about possible 
concerns when operating CGNs. The responding ISPs raised several concerns 
regarding the setup and the operation of CGNs.  A primary concern is that some 
applications (e.g., online gaming) do not work seamlessly with their CGN setups,
causing subscriber complaints that remain difficult for the ISPs to
resolve at the best of times.  Additional concerns relate to traceability of 
users behind CGNs.  Losing the ability to directly identify users can raise two 
kinds of problems.  First, ISPs may be legally required to be able to map flows 
to subscribers. Second, IP addresses accrue reputations as they get 
used---e.g., as sources of spam---and therefore by sharing IP addresses among 
users the reputation is also shared and can cause problems for some users.

In addition, operators voiced
concerns about a lack of well-developed best practices for configuring
and dimensioning CGNs, rendering operating these devices cumbersome.
In particular, operators need to resort to experimentation on aspects
such as the distribution of external IP addresses and port ranges to
customers, and whether to use distributed or centralized CGN
infrastructure.  Respondents named the port space as well as the
amount of state CGNs need to maintain as primary challenges when
configuring CGNs.  Accordingly, ISPs report widely varying
dimensioning of their CGNs in practice, ranging from static 1:1 NAT
per customer---to prepare for the future---to limits of 512~sessions
per customer due to heavy NATing.

\begin{figure}
\centering
      \includegraphics[width=.95\linewidth]{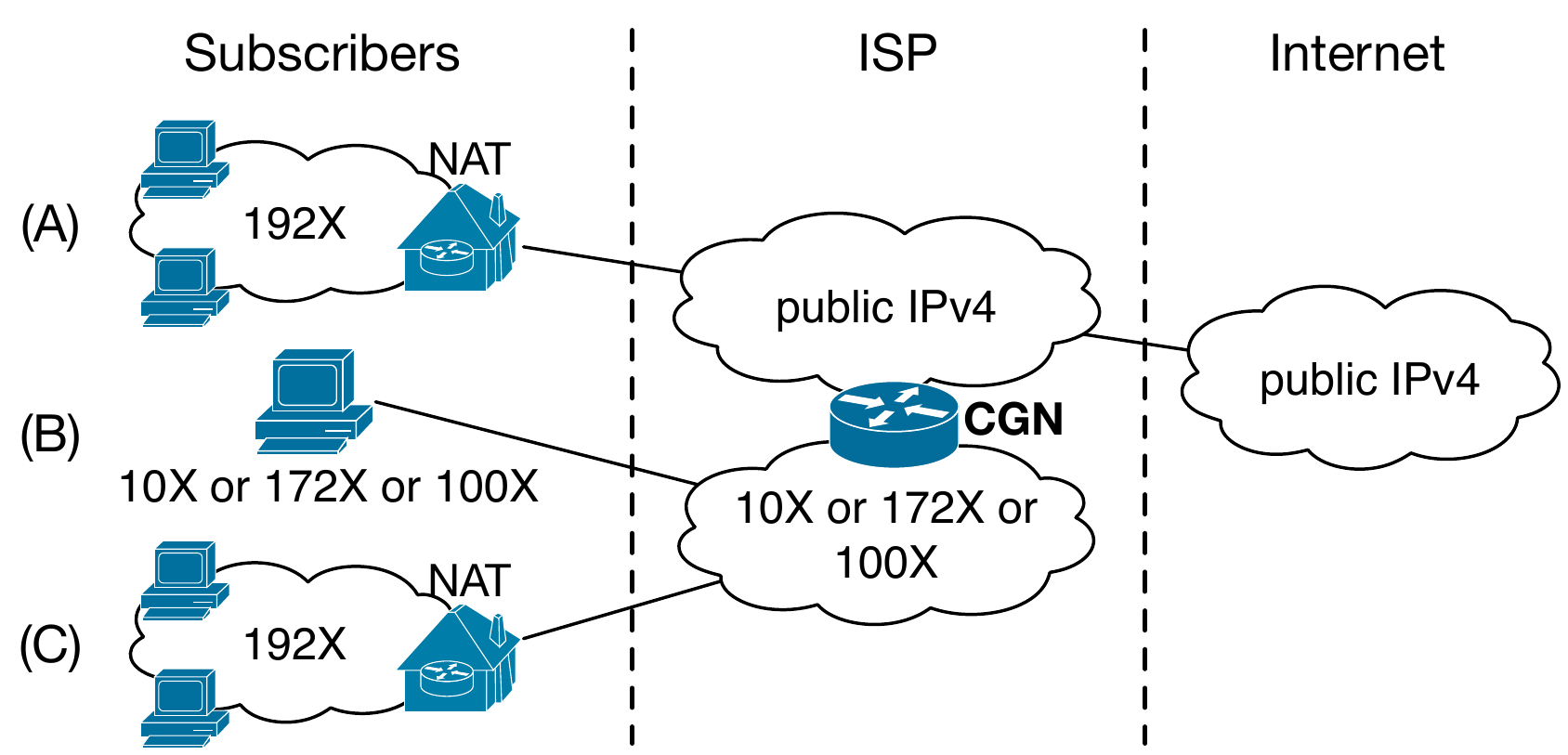}
      \caption{NAT scenarios. $A$ resides behind a single in-home NAT,
        $B$ behind a single carrier-grade NAT, and $C$ behind both an
        in-home and carrier-grade NAT (NAT444).}
  \label{fig:nat444_sketch}

\end{figure}

\section{Background}
\label{sec:background}

As we sketch above, the lack of ready access to new IPv4 addresses
is leading ISPs to alternate technologies to accommodate their
addressing needs. One such approach is to leverage Carrier-Grade
NATs (CGN). When an ISP uses CGN, it provides subscribers with internal IP 
addresses and then applies address translation to their traffic.  CGNs often 
introduce multiple layers of address translation since subscribers often run NAT
devices on their own edge networks (e.g., as built into most CPE
devices in users' homes). We refer to the case of subscribers whose
packets are translated once before they reach the public Internet as
\textit{NAT44} and to the case where packets are translated twice as
\textit{NAT444} \cite{nat44and444,address_sharing_ton}.

Figure~\ref{fig:nat444_sketch} illustrates various addressing
structures in common use on the Internet. In each of the scenarios the ISP
has a pool of public IPv4 addresses that are used differently by
various subsets of its customers.  The ISP gives each subscriber in
group \textit{A} a single public IP address.  The subscriber in
turn runs a subscriber-side NAT44 device to share this IP address
among all the devices on the internal network.  This is typical for
many residential subscribers.  Subscriber \textit{B} receives an
internal IP address from the ISP which a NAT translates into a public IP
before packets reach the wide-area network.  This case of a
carrier-side NAT44 device is common within cellular networks.
Finally, subscriber \textit{C}'s network is identical to subscriber
\textit{A}'s in that a local NAT is used to facilitate connectivity
for a multitude of internal devices.  However, in this case, instead
of providing a single external IP address the ISP provides the
subscriber with a single internal IP address, which in turn it
translates with a CGN before traffic reaches the wide-area network.
This is a case of NAT444, or two layers of address translation.
An ISP that runs a CGN does not necessarily NAT all of its subscribers. Many 
ISPs only NAT new subscribers and some even have various classes of subscribers 
and allow customers to choose their type of connectivity, which may come at 
different prices (some ISPs charge their customers for a public IP 
address, e.g., \cite{nowispnz}).

On the basis of the terminology used in the IETF, we now define several NAT-related terms we use throughout the
remainder of the paper.

\begin{table}[t]
\centering
\resizebox{\columnwidth}{!}{
\footnotesize
\begin{tabular}{llll}
\textbf{Range} & \textbf{Shorthand} & \textbf{RFC} & \textbf{Comments} \\ 
\hline
\TblSpB
192.168.0.0/16 & 192X & 1918 & Commonly used in CPE \\
172.16.0.0/12 & 172X & 1918 & \\
10.0.0.0/8 & 10X & 1918 & \\  
100.64.0.0/10 & 100X & 6598 & for CGN deployments \\ 
\end{tabular}
}
\caption{Address space reserved for internal use.}
\label{tab:privatespace}
\end{table}

\parax{Address Types}
We distinguish IP addresses both in terms of their location relative
to a NAT, as well as in terms of their numeric value.  We refer to an
address on the edge-facing, client-local side of a NAT as {\it
  internal} vs. {\it external} when nearer to the network core.  An
address is {\it reserved} if it resides in prefixes (as set forth
e.g., by RFC 1918 \cite{rfc1918}) that should not get announced to the global 
routing table, and {\it routable} otherwise. Table \ref{tab:privatespace} lists 
those address ranges reserved for internal use by the IETF. 
\footnote{Technically some reserved addresses are in fact routable; we
  focus here on their intended use.}

\parax{NAT Mappings}
NATs keep state that maps each internal IP address and
port number tuple to an external IP address and port
number tuple.  Unless manually configured, NATs create mappings
on-demand once a local host behind the NAT (\ie with an internal IP
address) sends a packet from its \privateipport endpoint to a
remote \dstipport. The NAT then records an \publicipport tuple,
translates the packet and sends it to the destination host. When the
external host replies to \publicipport, the NAT finds the
corresponding entry in its mapping table, translates the destination
address to \privateipport and forwards the packet internally.

\parax{Mapping Types}
NAT behavior differs in the reuse of existing mappings and in the 
filtering rules for the usage of established mappings. A 
\textit{symmetric NAT} creates different mappings for subsequent packets 
sent from the same \privateipport endpoint to different \dstipport 
endpoints. This behavior significantly impedes NAT traversal and makes 
symmetric NATs the most restrictive type of NAT. Other types of NAT 
reuse existing mappings regardless of their \dstipport. They differ 
in their filtering policy, here listed in decreasing order of restrictiveness: 
\textit{port-address restricted NATs} only allow incoming packets from the very 
\dstipport that was initially contacted from the host inside the NAT, 
\textit{address restricted NATs} require a matching \dstip, 
but allow packets from varying port numbers, while 
\textit{full cone NATs} allow incoming packets from any external host once a 
mapping is created. This makes full cone NATs the most permissive type of 
NAT~\cite{rfc3489}.\footnote{This terminology allows arranging NATs 
according to their restrictiveness and improves readability, therefore 
we use it despite being discouraged by the IETF~\cite{rfc4787}.}

\parax{Mapping Timeouts}
As with any stateful middlebox, NATs must manage their internal
state and therefore track active flows.  The NAT must release
mappings that are no longer needed.  NATs generally use both TCP
state tracking and timeouts to prune unnecessary NAT mappings.
Recommended minimum timeouts are 120~seconds for UDP~\cite{rfc4787}
and 2~hours for TCP~\cite{rfc5382}.

\parax{Port Allocation}
NATs differ in their selection of an external \publicport number for a
new session.  NATs implementing \textit{port preservation} attempt
to retain the original source port as the external port (i.e.,
\privateport$=$\publicport), unless there is a collision and an
alternate port must be chosen.  Other NATs---especially large
NATs---assign ranges of the external port space to each internal host
and then assign external ports on-demand from this pool in
sequential or random order~\cite{rfc4787}.

\parax{IP Pooling}
Large NATs typically use multiple external IP addresses, called
\textit{NAT pooling}. Upon connecting, a subscriber typically gets allocated a 
public IP address out of the pool. NATs employing \textit{paired pooling} always
use the same \publicip for a given \privateip.  Otherwise, a NAT is
said to use \textit{arbitrary pooling}. In our methodology, the presence of NAT 
pools will play an important role when it comes to dissecting home NAT 
deployments from CGN deployments.

\parax{Hairpinning}
Consider the communication between two hosts---$A$ and $B$---behind
the same NAT.  When $A$ sends a packet to $B$ it will use $B$'s
\publicipport.  When the NAT receives this packet it can detect that
the destination of the packet is in fact itself and therefore direct
the packet to $B$'s \privateipport.  This behavior is called
\textit{hairpinning} \cite{rfc4787,rfc5382}. If the NAT leaves the 
source \privateipport in place when forwarding the packet, then the hosts can 
discover their internal IP addresses when communicating behind the same NAT.

\subsection{Related Work}

IETF RFCs contain most of the available literature about CGNs. In particular, 
RFC 6888 specifies basic requirements for CGNs~\cite{rfc6888}, whereas
RFC 6544~\cite{rfc6544} and RFC 5128~\cite{rfc5128} describe two popular
mechanisms for NAT traversal: ICE and UDP/TCP hole punching, respectively.
As a result of NAT's added complexity, RFCs also
describe how CGNs affect application-level functionality~\cite{rfc5128,rfc7021}.

Several academic studies have tried to identify NAT deployment in home 
networks using UPnP queries~\cite{dicioccio2012probe,nat444pam}
or IP ID header fields~\cite{nat_detection_ipid:imw}, by passively
observing IP TTLs and HTTP User-Agent strings~\cite{residential_nat:pam}, and
by applying NAT detection to NetFlow traces~\cite{natdection_passive:conext}.
M{\"u}ller~\etal conducted an active, topology-based traversal of cascaded
large-scale NATs~\cite{muller2013analysis}. One NAT test presented in our work 
is an augmented version of their methodology. Ford~\etal 
studied the effectivity of different NAT punching techniques in NAT-ed 
networks~\cite{fordHolePunching}. 
The studies conducted by Wang~\etal~\cite{Wang:2011:USM:2043164.2018479} 
performed a comprehensive active measurement campaign to understand middleboxes
present in cellular networks. In contrast to Netalyzr their tool
relied on rooted handsets to modify packets at the IP and TCP layers. Donley 
\etal~\cite{cgnimpactwebbrowsing} studied the 
impact of CGN deployment on Web browsing performance in one ISP. Ohara 
\etal~\cite{ohara2014cgnmobile} simulated how CGNs can impact on TCP connection 
establishment in mobile networks. Finally, Skoberne \etal presented a 
theoretical taxonomy of NAT deployments and discuss their possible impact on 
network performance~\cite{address_sharing_ton}.  
Richter~\etal~\cite{IMC2016Beyond} measured an increasing concentration of 
traffic on fewer IPv4 addresses during 2015, hinting at an
increasing use of CGN deployment in the Internet.

Little is known about actual CGN hardware deployed in the wild and their 
consequences for the different stakeholders. We cannot readily identify how NAT 
vendors implement their equipment and how ISPs take advantage of them. To 
partly overcome this limitation, we rely on vendor manuals and network operator 
tutorials to obtain deeper insights into practical considerations of CGN
deployment~\cite{cisconat,a10nat,nat44and444,cgn_menog}.

\section{Detecting CGN at Scale}
\label{sec:cgndetection}

Our first set of methodologies aim to investigate the breadth of CGN
deployment in the Internet.  In general terms our CGN detection
mechanisms leverage both internal and external observations of IP
addresses associated with a given host to detect discrepancies and
therefore presence of address translation.  We use two techniques to
obtain internal observations: implicit and explicit.  Our implicit
observations come via standard BitTorrent clients leaking internal
address information, while our explicit observations come from users
running active measurements on our behalf via our Netalyzr tool.  We
stress that we strive for conservativeness in our CGN detection
methods.  That is, we would rather provide a sound lower bound on
CGN presence than using a more speculative approach that
identifies more CGNs of questionable validity.

\subsection{Detecting CGNs via BitTorrent}
\label{sec:cgndetection:bt}

The BitTorrent Distributed Hash Table (DHT) \cite{bittorrent_dht}
represents a distributed data structure that links hosts looking for
specific content with hosts that have that content without using
centralized infrastructure.  The nodes that make up the DHT form a
connected graph so that search queries for specific content are
propagated to a node with the given information.  Each node is
identified by a 160~bit \btnodeid which is randomly chosen by the
node itself (and is unique with high probability).  To form the
graph, DHT nodes both maintain a list of DHT peers and provide an
interface for other nodes to query this list.  Further, the nodes on
this list must be periodically validated with \btping messages to
ensure reachability.  This in turn means that the contact
information a node $A$ has for node $B$---in the form of an IP
address and port number---represents $B$'s location from $A$'s
perspective.  We observe that the nodes represent vantage points
that we do not control but can none-the-less probe to learn about
host-to-host connectivity.  We find that this connectivity is
sometimes represented by internal IP addresses.  That is, the path
between two hosts does not traverse the publicly routed Internet,
but takes place completely within a private network (e.g., within an
ISP).  Additionally, these hosts are clearly also able to
communicate outside of this private network and therefore are behind
some form of NAT.  We developed a crawler to collect connectivity
information from the BitTorrent DHT and then leverage that data to
form an understanding of CGN deployments.

\parax{Crawling the BitTorrent DHT}
We developed a crawler that starts with DHT nodes learned from the
BitTorrent bootstrap servers and issues a series of \btfindnodes
requests to DHT nodes with a random query.  The response to
\btfindnodes is a list of up to eight ``close'' peers where
closeness is calculated using the XOR distance between the query and
each \btnodeid in the node's list of peers
\cite{maymounkov2002kademlia}.  We issue five queries, which provides
connectivity information---\btnodeid, IP address and port
number---for roughly 40~nodes.  We then in turn query the newly
learned peers in the same fashion.  Our crawler also participates in
the DHT and therefore accepts incoming requests from nodes that have
learned about our crawler through the source information in our
requests.

\begin{table}
  \centering
  \resizebox{0.78\columnwidth}{!}{
  \footnotesize
  \begin{tabular}{p{2cm} r r r}
    {}& {\bf Peers} & {\bf Unique IPs} & {\bf ASes} \\ 
\hline
\TblSpB
{\bf Queried} & 21.5M~ & 15.5M~ & 18.8K~ \\
{\bf Learned} & 192.0M~ & 62.1M~ & 26.7K~ \\
\end{tabular}
}
  \caption{BitTorrent DHT data: \textit{Queried}: Peers that were issued and 
  replied to \btfindnodes requests. \textit{Learned}: All peer information we 
  gathered.} 
  \label{tab:bt_crawler_generalstats}
\end{table}

\begin{table}
  \begin{center}
  \resizebox{\columnwidth}{!}{
  \small
\begin{tabular}{p{1.1cm} r r | r r r}

& \multicolumn{2}{c}{\textbf{\footnotesize Internal Peers}}
& \multicolumn{3}{c}{\textbf{\footnotesize Leaking Peers}} \\

{\bf Range}& {\bf Total} & {\bf Unique IPs} & {\bf Total}  & {\bf Unique IPs} & 
{\bf ASes} \\
\hline
\TblSpB
{\bf 192X} & 565.9K~ & 11.2K~ &
186.8K~ & 162.2K~ &
4.1K~ \\
{\bf 172X} & 336.6K~ & 85.0K~
& 52.9K~ & 33.9K~ &
1.0K~ \\
{\bf 10X} & 1.3M~ & 328.5K~ &
283.9K~ & 194.4K~ &
2.2K~ \\
{\bf 100X} & 1.5M~ & 251.5K~ &
192.0K~ & 165.8K~ &
723~ \\
\end{tabular}
}
  \caption{Peers reported via reserved IP addresses (left) and
  the corresponding peers that leaked them (right).}
      \label{tab:bt_private_ips}
  \end{center}
    \vspace{-1em}
\end{table}

As we note above, in some instances peers reply to \btfindnodes with
information about nodes that include reserved IP addresses
(Table~\ref{tab:privatespace}), indicating the probed peer can reach the 
reported peer without crossing the publicly routed Internet.  We refer to this 
behavior as \textit{internal address leakage}.  When we learn an internal 
address for a given \btnodeid we refer to this node as an \textit{internal 
peer}.  When our crawler finds a node leaking internal peers we issue an 
additional ten \btfindnodes queries in the hopes of finding additional internal 
peers.  We continue issuing \btfindnodes queries in batches of ten for as long
as we continue to harvest internal peers.

Note that within BitTorrent the \btnodeid is the sole identity notion for a 
given peer.  However, as peers can have multiple endpoints (internal, 
external), as well as multiple IP addresses/ports due to dynamic IP address 
assignment or due to BitTorrent clients modifying the local port number, we 
identify a unique peer by the full tuple of \textit{(IPaddress:port, nodeid)}.
As a positive side effect, this also eliminates possible biases due
to DHT poisoning \cite{wang2012real}, where peers announce themselves with 
a foreign \btnodeid.

The dataset we use in the remainder of this paper comes from a one-week crawl 
starting on March 3, 2016.\footnote{We have additional crawls from late 2015 
and early 2016 that show  consistent results to those we present in this paper.}
Table~\ref{tab:bt_crawler_generalstats} summarizes the dataset.  We
probed more than 21M peers across nearly 19K ASes.  These probed DHT
nodes in turn revealed contact information for 192M peers across
more than 26K ASes.  Of these 192M peers, 107.7M~
peers and 36.7M~ unique IP addresses responded to
\btping probes. Table \ref{tab:bt_private_ips} shows an overview of the leaked 
contact information, where we break private peers down based to the internal 
address space range. Among the peers crawled, we find more than 700K peers 
leaking contact information for more than 3.7M internal peers (i.e., peers with 
IP addresses in a reserved range) across more than 5K ASes.

\parax{Identifying CGNs}
Our dataset clearly shows the presence of NATs via leaked internal
peers.  Next we seek to establish the degree to which these NATs are
network-level CGNs as opposed to simple home NAT deployments.
First, to detect any type of NAT using the BitTorrent dataset there
must be multiple BitTorrent clients that directly communicate within
some internal network.  Next, to determine the presence of a large
network-level CGN we require \textit{NAT pooling} behavior
(\xref{sec:background}).  In other words, within a single AS we
require \first multiple peers with different external IP addresses
to leak internal peers and \second intersections in the internal
peers leaked across multiple external IP addresses. Moreover, we require the 
internal peers within a cluster to reside within the same internal address 
range (e.g., 10/8).

\begin{figure}
  \begin{center}
  \subfigure[Isolated leaking relationships (AS7922, Comcast, 192X internal 
  space).]{
    \includegraphics[width=0.45\linewidth]{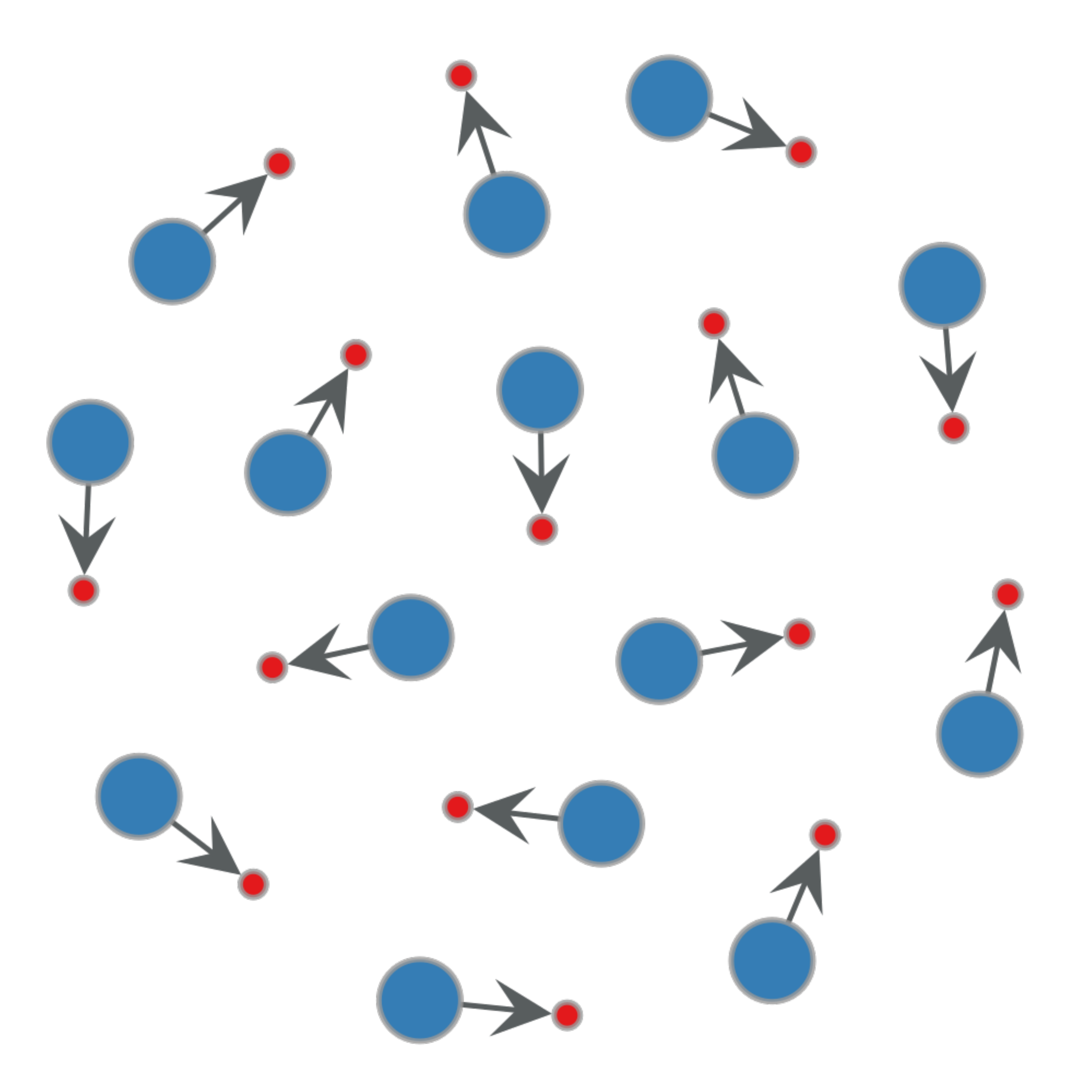}
    \label{fig:privatepeer_clustering_nocluster}
  }
  \hfill
  \subfigure[Clustered leaking relationships (AS12874, FastWEB, 100X internal 
  space).]{
    \includegraphics[width=0.45\linewidth]{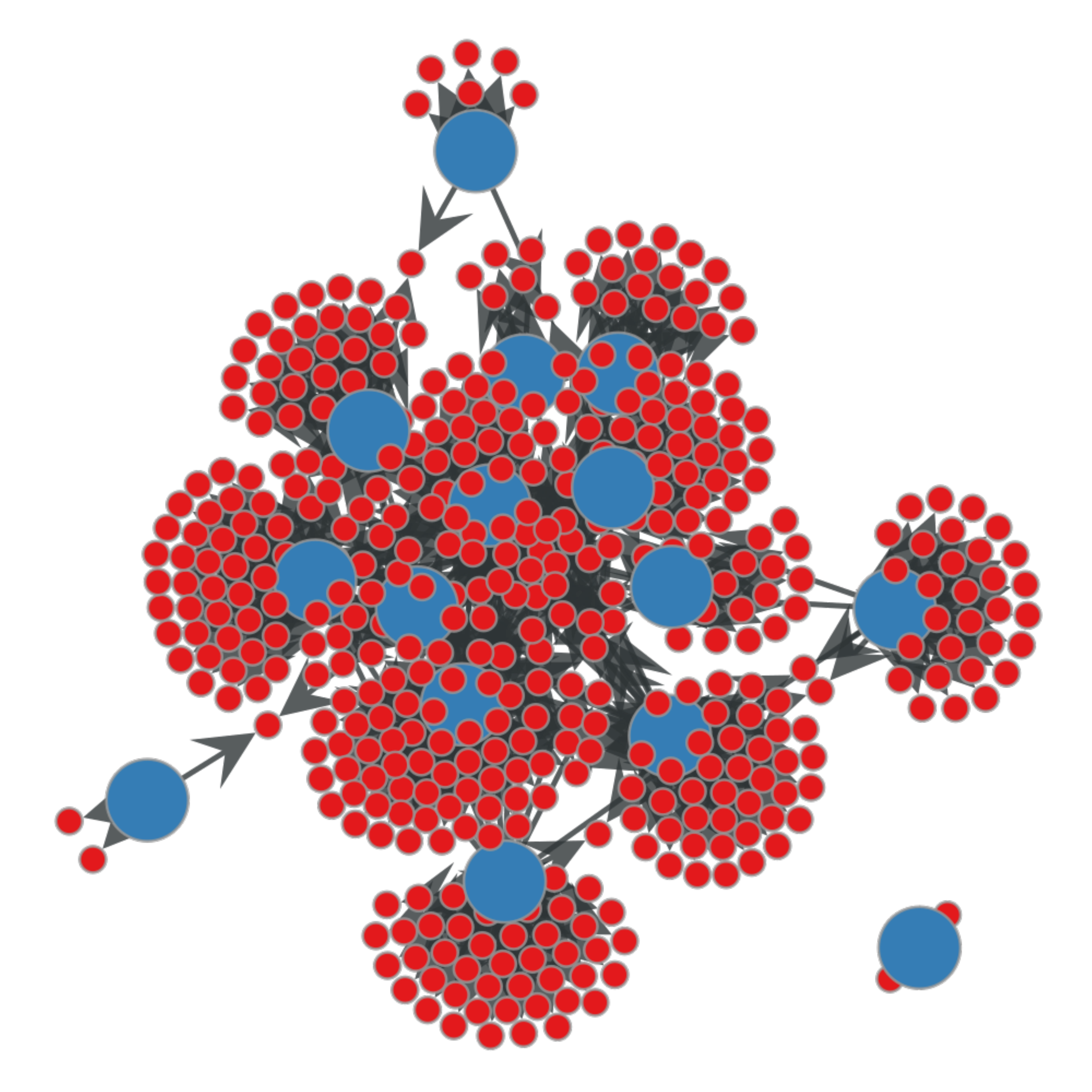}
    \label{fig:privatepeer_clustering_bigcluster}
  }
  \caption{Peer leakage in non-CGN vs. CGN ASes. Large blue vertices are 
  BitTorrent peers leaking peers with internal IP addresses (small red 
  vertices).}
   \label{fig:privatepeer_clustering}
  \end{center}
    \vspace{-1em}
\end{figure}

To detect this behavior on a per-AS level we next form a graph for
each AS where each peer is a vertex and each edge between a public
peer $A$ and an internal peer $B$ indicates that $A$ leaks contact
information for $B$.  Note, when constructing graphs we only
consider internal peers which were leaked exclusively by peers
residing in a single AS.  This excludes leaking relationships caused
by VPN tunnels.  Figure \ref{fig:privatepeer_clustering} shows a
small subset of the graphs for two 
ASes as an illustration.
Figure~\ref{fig:privatepeer_clustering_nocluster} shows there is
only isolated leaking within AS~7922 (Comcast).  We find more 
than 1K peers leaking internal addresses within AS~7922.  However,
we also find that each leaked internal peer is leaked by exactly one
external peer.
In contrast, Figure~\ref{fig:privatepeer_clustering_bigcluster} shows strong 
clusters within AS~12874 (FastWEB) consisting of multiple peers behind 
different external IP addresses that leak a large number of internal peers, 
which form large intersections.\footnote{We manually confirmed CGN presence 
in AS~12874 and also verified the discovery of internal peers (via \textit{NAT 
Hairpinning}, \xref{sec:background}) and leakage of internal peers in 
this AS by running a regular BitTorrent client on a host behind CGN in this 
AS.} This shows that our clustering methodology is effective in 
separating home NAT deployments from network-level CGNs.

\begin{figure}
   \begin{center}
     \includegraphics[width=0.95\linewidth]{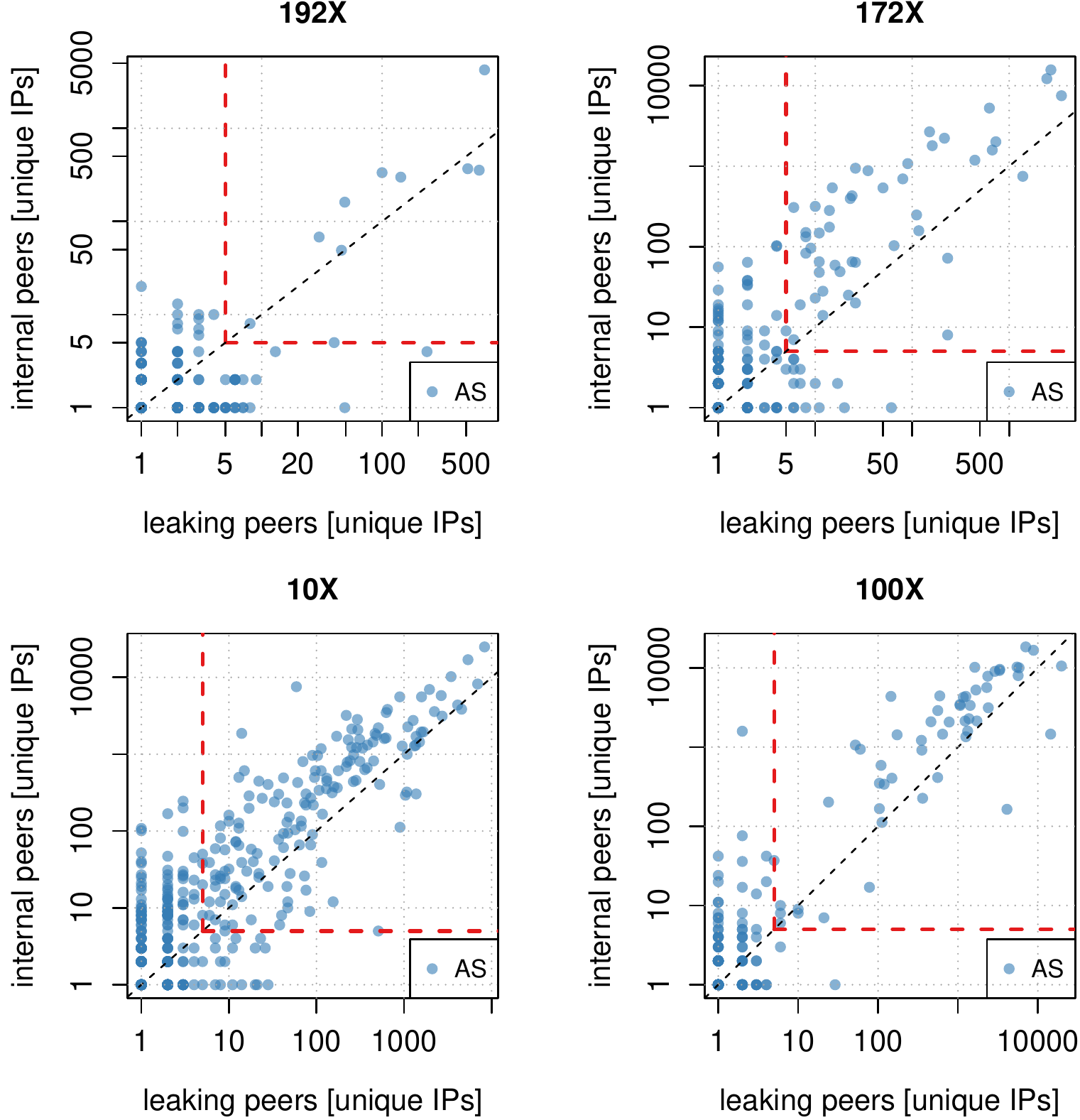}
   \caption{Size of the largest connected cluster of leaking and internal 
   BitTorrent peers per AS. The x-axis shows the number of public IP addresses, 
   the y-axis the number of internal IP addresses contained in the largest 
   cluster.}
     \label{fig:btnatclustering}
   \end{center}
     \vspace{-1.5em}
 \end{figure}

We next construct a graph for each AS in our dataset and determine the largest 
connected cluster for each AS. Figure~\ref{fig:btnatclustering} shows our 
clustering results. Here, we plot a point for each AS and position it according 
to the size of the largest cluster we found in this AS (if any). In particular, 
the \textit{x}-axis shows the number of unique public IP addresses contained in 
the 
largest cluster and the \textit{y}-axis shows the number of unique internal IP 
addresses contained in the largest cluster. 
We only find a small number of ASes that contain large clusters in
the 192X space  
(top left figure).  We find ASes with large clusters to be more
prevalent in the 
other, larger, internal address ranges. This supports our hypothesis
that 192X address space  
is primarily used in small home NAT environments.
While in principle a cluster with 
at least two different external IP addresses is indicative of \textit{NAT 
pooling}, we only determine CGN presence for an AS when the largest connected 
cluster contains at least five public IP addresses and five private IP 
addresses. This is to address possible misclassifications arising from dynamic 
addressing, e.g., a home network with internal NAT deployment that changes its 
public IP address. We annotated Figure~\ref{fig:btnatclustering} with our 
detection boundary.
While we show network-wide results in \xref{sec:networkwideview}, we note that 
this methodology shows CGN usage in roughly 10\% of the probed ASes
for which our crawler queries at least 200~peers. 

\parax{DHT Data Calibration} Our BitTorrent-based CGN detection
relies on three key properties of the DHT peers: \textit{(i)}
BitTorrent peers behind the same NAT can learn internal endpoints of
other peers, \textit{(ii)} peers export internal endpoints via the
DHT, and \textit{(iii)} peers only propagate contact information for
peers that have been validated via direct interaction.  We verified
\textit{(i)} and \textit{(ii)} by running two popular and unmodified
BitTorrent clients (uTorrent on Windows and Transmission on Linux)
and measuring the control traffic they exchange as part of their
regular operation. We confirmed that these peers learned their
internal endpoints when located behind NATs that allow multicast
communication as well as behind NATs that have \textit{Hairpinning}
enabled.  Further, these peers forward packets with internal source
IP addresses (see \xref{sec:background}). We also validated the
latter within an ISP that deploys CGN. Therefore, we conclude that
BitTorrent clients can---if the circumstances allow it---learn their
internal endpoints and propagate that information via the DHT when
requested.

Finally, we assume hosts follow the BitTorrent DHT specification
\cite{bittorrent_dht} and only propagate reachability information for peers they
learn after reachability has been directly validated by the host
itself.  Otherwise, hosts would propagate potentially dubious
reachability information and likely we would detect 
CGN presence in practically any AS that hosts enough peers.  To
validate our assumption, we setup a common BitTorrent client  
(uTorrent on Windows) with a \btnodeid of $ID_{us}$ and let it
interact with the DHT.  At the same time on a different host we
crawled the DHT requesting $ID_{us}$.  We queried 100K peers and
were given contact information for $ID_{us}$ by $1,387$ peers.  We
found that only 18 of these DHT peers (1.3\%) did \textit{not}
validate the reachability of $ID_{us}$ before propagating the
information.  This shows that our assumption that DHT peers follow
the specification and properly validate reachability before
propagating contact information is sound.

\subsection{Detecting CGNs via Netalyzr}
\label{sec:cgndetection:neta}

To complement our observations from crawling the BitTorrent DHT, we
leverage ICSI's \neta network troubleshooting service
\cite{netalyzr:imc:short}.  While the BitTorrent DHT provides a
useful set of specific information from end hosts, \neta allows us
to define explicit tests we wish to run from end hosts.  These tests
interact with a suite of custom-built test and measurement servers
and return the results to our data collection server.  We developed
a set of tests aimed at illuminating NAT behavior and deployed these
in 2014.  While \neta provides the potential to gather much richer
information than we find in the BitTorrent DHT, we are at the mercy
of individual users to access \neta and run the tests.  Users run
\neta via one of three supported clients: a Java applet for Web
browsers, a command-line client, or an Android client available in
the Google Play store~\cite{netalyzrGPlay}.

\begin{table}
\resizebox{\columnwidth}{!}{%
  \small
\begin{tabular}{p{2.7cm} | r | r r} 
& \textbf{Cellular} & \multicolumn{2}{c}{\textbf{Non-cellular}} \\
& {\bf $IP_{dev}$} & {\bf $IP_{dev}$} & {\bf $IP_{cpe}$} 
\\
{\bf Address Space} & $N$=8.6K~ & $N$=567.5K~ & 
$N$=229.8K~ \\
\hline
\TblSpB
\textbf{192X} & 0.2\%~ & 
92.4\%~ & 8.9\%~ \\
\textbf{172X} & 2.5\%~ & 
1.1\%~ & 0.8\%~ \\
\textbf{10X} & 58.7\%~ & 
6.2\%~ & 4.8\%~ \\
\textbf{100X} & 17.3\%~ & 
0.0\%~ & 1.9\%~ \\

\textbf{unrouted} & 12.5\%~ & 
0.0\%~ & 0.0\%~ \\
\textbf{routed match} & 5.7\%~ & 
0.0\%~ & 83.0\%~ \\
\textbf{routed mismatch} & 3.0\%~ & 
0.3\%~ & 0.5\%~ \\
\end{tabular}
}
  \caption{Address ranges seen for the device IP address and for the router's 
  external IP address.} 
      \label{tab:netalyzr_ips}
  \vspace{-1.2em}
\end{table}

In the context of understanding CGNs, \neta offers two advantages
over our BitTorrent crawl.  First, since BitTorrent is not heavily
used on mobile devices the Android version of \neta extends our view
into this important network type.\footnote{Note, while mobile
  devices can join wifi networks we scope our measurements to those
  on cellular data networks.}  Second, \neta allows us to directly
obtain the IP addresses used by the host, including $(i)$ the local
IP address of the device that executes \neta, $IP_{dev}$, $(ii)$ the
external IP address of the CPE router device as learned via UPnP
(where available), $IP_{cpe}$, and $(iii)$ the public IP address as
seen by our Netalyzr server, $IP_{pub}$.  We categorize $IP_{dev}$
and $IP_{cpe}$ in four categories: $(i)$ private address from one of
the reserved blocks for this purpose, $(ii)$ \textit{unrouted} for
addresses that are nominally public, but do not appear in the
routing table, $(iii)$ \textit{routed match} for case where the
address is routable, appears in the routing table and matches
$IP_{pub}$ (i.e., the non-NAT case) and $(iv)$ \textit{routed
  mismatch} for the case where the address is routable, appears in
the routing table but does not match $IP_{pub}$.

  \begin{table*}
  \begin{center}
  \resizebox{\textwidth}{!}{%
  \begin{tabular}{p{4cm} | r r | r r | r r} 
  
  & \multicolumn{2}{c|}{\textbf{routed ASes (N=52K)}} 
  &\multicolumn{2}{c|}{\textbf{eyeball 
  ASes, PBL (N=2.9K)}} & \multicolumn{2}{c}{\textbf{eyeball 
  ASes, APNIC (N=3.1K)}} \\
  
  &  covered & CGN-positive & covered & CGN-positive & covered & CGN-positive \\
  \hline
  \TblSpB
  \textbf{BitTorrent} & 
  2,724 (5.2\%) & 254 (9.40\%) & %
  1,673 (57.7\%) & 180 (10.8\%) & %
  1,824 (59.6\%) & 204 (11.2\%) \\ %
  \textbf{Netalyzr non-cellular} & 
  1,367 (2.6\%) & 195 (14.3\%) & 
  866 (29.8\%) & 151 (17.4\%) & 
  929 (30.4\%) & 174 (18.7\%) \\
  \textbf{BitTorrent $\cup$ Netalyzr} & 
  3,166 (6.0\%) & 421 (13.3\%) &
  1,791 (61.7\%) & 306 (17.1\%) & 
  1,946 (63.6\%) & 350 (18.0\%) \\
  \hline
  \TblSpB
  \textbf{Netalyzr cellular} & 
  218 (0.4\%) & 205 (94.0\%) &
  175 (6.0\%) & 162 (92.6\%) & 
  171 (5.6\%) & 161 (94.2\%) \\
  \bottomrule
  \end{tabular}
  }
    \caption{Coverage and detection rates of our methods as fraction of all 
    routed ASes, as well as Eyeball ASes, primarily connecting end users, as 
    derived from PBL and APNIC.} 
        \label{tab:cgn_coverage_detection}
  \end{center}
    \vspace{-1em}
  \end{table*}

\parax{Cellular Networks}
Detecting the use of CGNs in cellular networks is straightforward in
\neta because there are no devices between the mobile device and the
ISP and therefore the classification of the ISP-assigned $IP_{dev}$
directly indicates the presence of a CGN.\footnote{Exceptions could
  be caused by users manipulating their network access with VPN
  tunnels or by users who run their own cellular access point (\eg
  femtocells). \neta's Android client collects enough data to allow
  us to prune such cases from our analysis.}  While straightforward,
we require five observations from an AS before we include it in our
study to ensure our conclusions are sound and are not the result of
some unexpected behavior.  The second column of
Table~\ref{tab:netalyzr_ips} shows the breakdown of $IP_{dev}$ for
all our cellular \neta sessions.  We find 94\% of the
sessions---i.e., all cases except the \textit{routed match}
case---show a translated address.
A first view of CGN deployment in cellular networks on a per-AS
basis shows 63.8\%~ exclusively
assign internal IP addresses to mobile devices.  Similarly, we find
6.0\%~ of ASes exclusively assign 
public IP addresses to devices and show no signs of address
translation.  Meanwhile,
30.3\%~ assign a mixture of
internal IP address and public IP addresses to devices. We note that
another 5.0\%~ of ASes assign
internal addresses from public blocks that are actually routed, but
still perform address translation!

\parax{Non-Cellular Networks}
We next shift to \neta's detection of CGNs in non-cellular networks.
In these networks, $IP_{dev}$ is often assigned by a device in the
device's network and therefore when CGNs are present there are
multiple address translations happening in the path (NAT444).  This in turn
makes detection difficult.  First, we winnow our analysis to ASes
that have at least ten \neta sessions in our dataset.\footnote{Note,
  we require more observations in the non-cellular case (ten)
  compared to the cellular case (five) because the situation is not
  as straightforward due to the presence of in-path network
  equipment in the edge network.  This makes the breadth of behavior
  we observe larger and in turn we need more observations to draw
  sound conclusions.} The third column of
Table~\ref{tab:netalyzr_ips} shows that $IP_{dev}$ is nearly always a
private address, as expected.  In addition to $IP_{dev}$, \neta uses
UPnP \cite{rfc6970} to attempt to determine $IP_{cpe}$ for the first
hop CPE device.
The fourth column of Table~\ref{tab:netalyzr_ips}
shows the breakdown of $IP_{cpe}$ for the 40\% of cases where UPnP
provides the address.  In 83\% of the cases, $IP_{cpe}$ is a public
IP address from the ISP, hence no CGN is present.  The remaining
17\% of the cases clearly point to multiple NATs.  However, whether
these are ISP-based CGNs or multiple small-scale NATs in the edge
networks is not clear.  Therefore, we add two steps to disambiguate
the situation.

 \begin{figure}
   \begin{center}
     \includegraphics[width=0.95\linewidth]{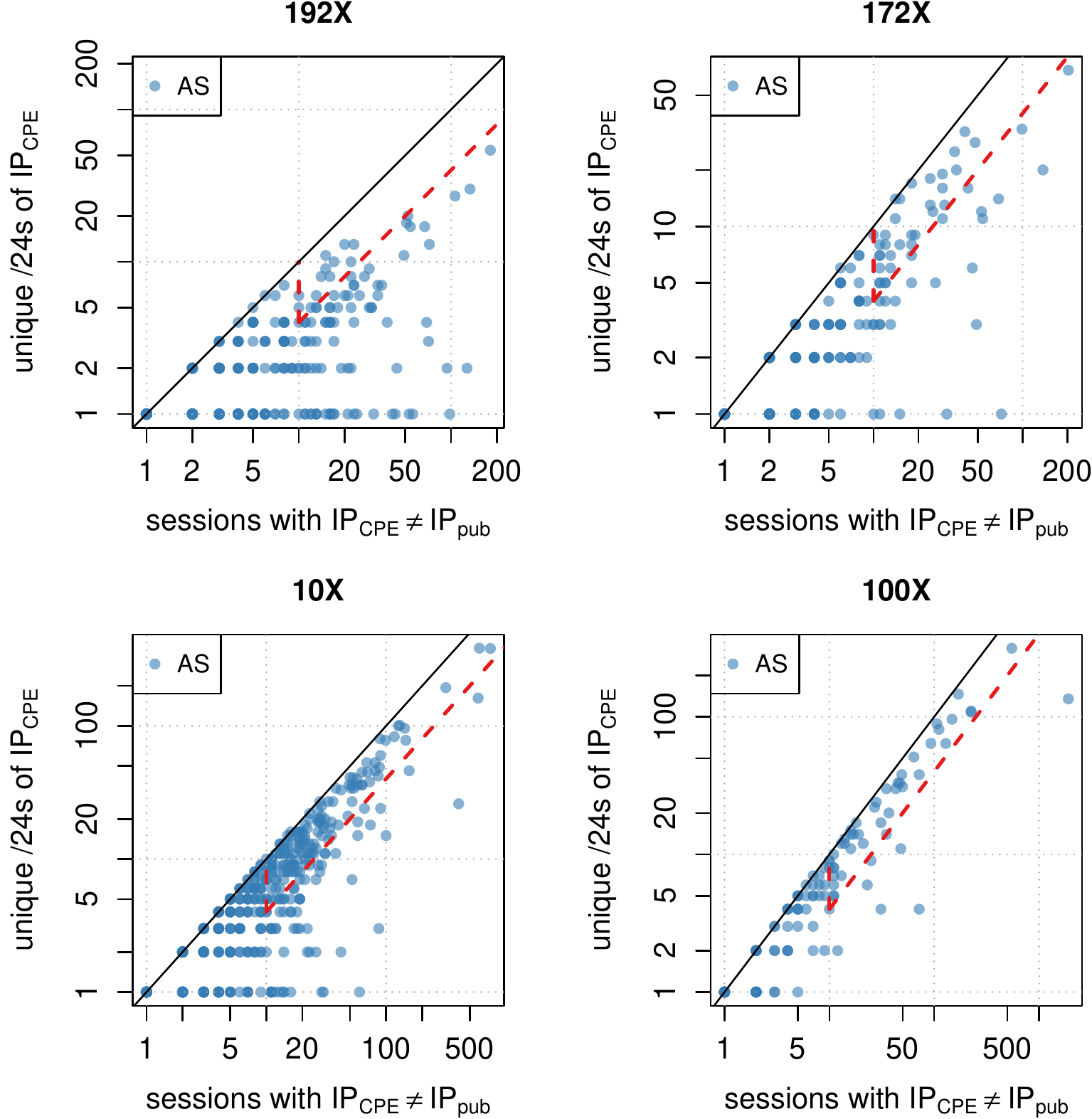}
   \caption{Netalyzr CGN candidate ASes: Sessions where $IP_{cpe}$ does not 
   match $IP_{pub}$ (x-axis) vs. unique /24s of $IP_{cpe}$ addresses 
   (y-axis).}
     \label{fig:upnpasscatter}
   \end{center}
   \vspace{-1em}
 \end{figure}

First, we observe that CPE routers often make assignments from the
192X block (Table~\ref{tab:netalyzr_ips}, column 3), whereas the
CGNs we find via BitTorrent and in the cellular environment more
often make assignments outside the 192X block
(Figure~\ref{fig:btnatclustering} and
Table~\ref{tab:netalyzr_ips}, column 2).  Therefore, we use \neta's
list of $IP_{dev}$ assignments to determine the top ten
/24 blocks from which CPE devices make assignments (covering 95\% of
assignments).  We then conclude that any $IP_{cpe}$ that falls
within one of these blocks was likely assigned by another local CPE
device and not a CGN.  Applying this filter removes more than half
the ambiguous situations and leaves us with 7.9\% of \neta's
sessions that may be CGNs.

As a second step, we observe that due to their scale, CGNs
necessarily must more broadly assign addresses than would be
necessary in a small-scale edge network.  Therefore, to conclude a
CGN is present we require $IP_{cpe}$ diversity within an AS.
Specifically, an AS must have $N~\ge~10$ \neta sessions that may be
behind a CGN.  We expect that as the number of \neta sessions
increases, our observations of address space diversity will, as
well.  Therefore, those sessions must span at least $0.4~\times~N$
internal /24 address blocks are deemed to indicate a CGN is
deployed.\footnote{We note that we do not expect address diversity
  to infinitely scale with the number of observations.  However,
  given our data this heuristic works well.  Furthermore, adding additional
  complexity to the methodology without grounding in empirical
  observation is not useful.}  Figure~\ref{fig:upnpasscatter} shows
a point for each AS in our dataset, with the $x$-axis showing the
number of ambiguous multiple NATing situations we observe and the
$y$-axis showing the number of /24 address blocks we observe within
the AS.  The dashed line represents our CGN detection cutoff point.
Similar to our observations in our BitTorrent dataset, the 192X
address space is sparsely used for CGNs, while more CGNs are present
in the other reserved address blocks.  Overall, our method detects
CGN presence in almost 15\% of the covered ASes.

Our CGN detection is no doubt imperfect.  However, we note that our
heuristics start with cases where our data \textit{conclusively}
indicates multiple address translators are present.  Further, manual
validation against our survey results, ISP' websites and threads on
operator mailing lists lends confidence to our conclusions.
Finally, as we note above, our methods for labeling CGNs are
conservative.  For instance, there are points to the right of dashed
line in Figure~\ref{fig:upnpasscatter} that likely represent
undetected CGNs.  These points represent many \neta sessions that
show much address diversity---but not enough to meet our threshold.
Our validations and conservative cutoffs leave us confident in the
determinations we make, at the likely expense of not identifying all
CGN deployments.

\section{A Network-wide view of CGN Deployment}
\label{sec:networkwideview}

We now summarize our measurements of global CGN deployment based on
the methodologies we develop in \xref{sec:cgndetection}.  Table
\ref{tab:cgn_coverage_detection} reports our results in terms of
ASes where we detect at least partial CGN deployment.  We report our
results within the context of three different AS populations in the
three big columns in the table.  The second big column of the table
considers the entire population of roughly 52K routed ASes.  Meanwhile,
the third and fourth columns represent the results in the context of
so-called ``eyeball'' ASes (ASes that connect end users to the Internet).
The third big column considers the
population of ASes that the Spamhaus Policy Block List \cite{pbl}
identifies as including the equivalent of at least 2,048~addresses
in ``end user'' blocks.  The last big column considers the
population of ASes to be those in the APNIC Labs AS Population
list \cite{aspop_apnic} that have at least 1,000 samples.
Our datasets cover 6.0\% of the ASes in the Internet, but over 60\%
of the eyeball ASes.  Given that our methodologies rely on
user-driven tools (Netalyzr and BitTorrent clients) it is
unsurprising that we cover an order of magnitude more eyeball ASes.

In terms of CGN deployments, we find that 13.3\% of all non-cellular ASes
use CGNs.  However, the penetration jumps to 17--18\% when
considering only non-cellular eyeball ASes.  In cellular networks
the use is over 92\% in all cases.  These results show that CGNs are
a reality for many Internet users.
We also note that while we are able to cover roughly twice as many
ASes with our BitTorrent dataset, the Netalyzr measurements find
CGNs in higher proportions.  This is expected and underscores
important aspects of each methodology.  While we are able to
opportunistically leverage the information from the BitTorrent DHT,
we are unable to direct or control the measurements.  So, while
BitTorrent has a large footprint the data is noisy.  On the other
hand, \neta must coax people to explicitly run the tool and
therefore the population is not large.  However, once run we
directly control the measurements and can gather more data directly
(e.g., via probing UPnP).  Finally, we note that the table shows
that \neta often does not add significantly to the coverage, but
does add significantly to the CGN deployment results.  Therefore,
the BitTorrent detection should be viewed as a lower bound on CGN
penetration.

\begin{figure}[t]
    \centering
    \includegraphics[width=\columnwidth]{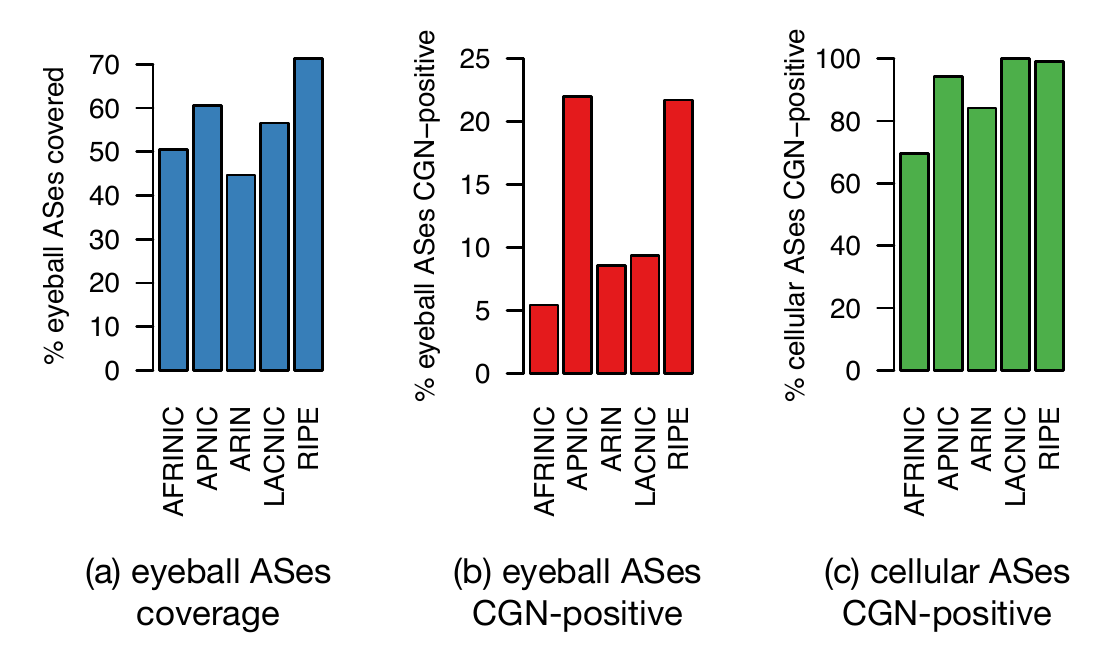}
    \caption{Eyeball AS coverage and CGNs per region.}
    \vspace{-1em}
    \label{fig:cgn_per_region}
\end{figure}

Finally, we return to the impetus of NAT in the first place: address
scarcity.  Figure~\ref{fig:cgn_per_region} shows our results by 
RIR.\footnote{The management of the address space is distributed over 5 
Regional Internet Registries (RIRs). ARIN serves the North American region, 
LACNIC the Central- and South American region, RIPE the European region, 
AFRINIC the African region, and APNIC the Asian region. For more information, 
we refer to \cite{RABP14}.}
The left-hand plot shows that the percentage of covered eyeball ASes
within each region does not show a significant regional
bias.\footnote{We use the PBL eyeball AS list for this plot.}  The
middle plot in the figure shows the percentage of the eyeball ASes
we find to deploy CGNs.  Here we observe that APNIC and RIPE show
more than twice the CGN penetration of the other regions.  These are
also the two regions that ran out of IPv4 addresses first.
Meanwhile, we find the lowest CGN penetration in AFRINIC, which is
the only region that has not yet exhausted its supply of IPv4
addresses.  The last plot in the figure shows the CGN penetration in
cellular networks by region.  AFRINIC is again an outlier in this
plot with ``only'' two-thirds of the ASes leveraging CGNs.

\begin{figure}
  \begin{center}
  \subfigure[Internal address space usage per CGN deployment.]{
    \includegraphics[width=0.95\linewidth]{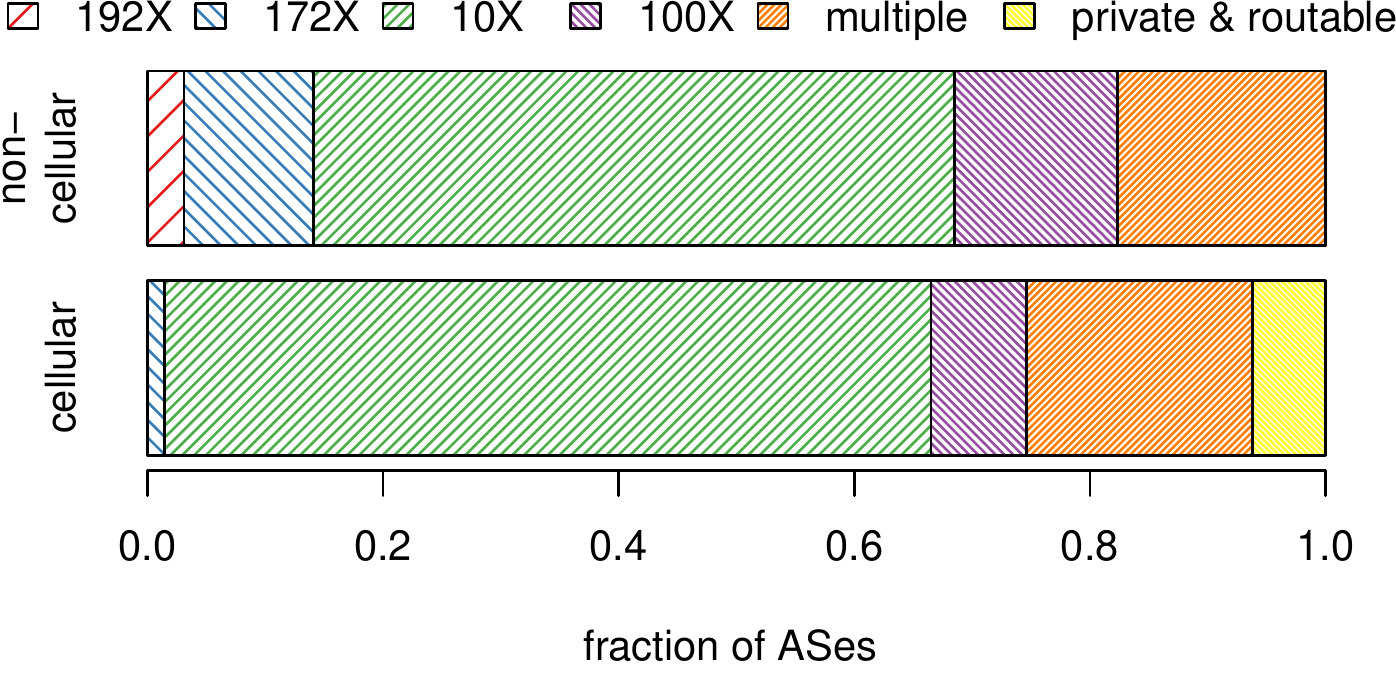}
    \label{fig:cgnaddresspacebar}
  }
  
  \subfigure[ASes using routable space as internal space.]{
    \includegraphics[width=0.80\linewidth]{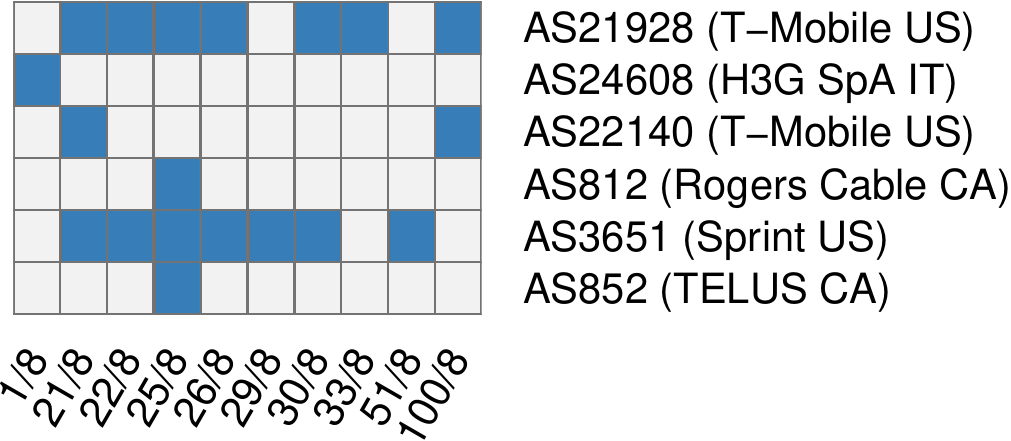}
    \label{fig:routablespaceases}
  }

  \caption{Internal address space in CGN deployments.}
  \label{fig:cgnaddressspaces}
  \end{center}
\end{figure}

\begin{figure*}
  \subfigure[Ephemeral port space seen by our server from non-CGN vs. CGN 
  connections without port preservation.]{
    \centering
    \includegraphics[width=0.3\linewidth]{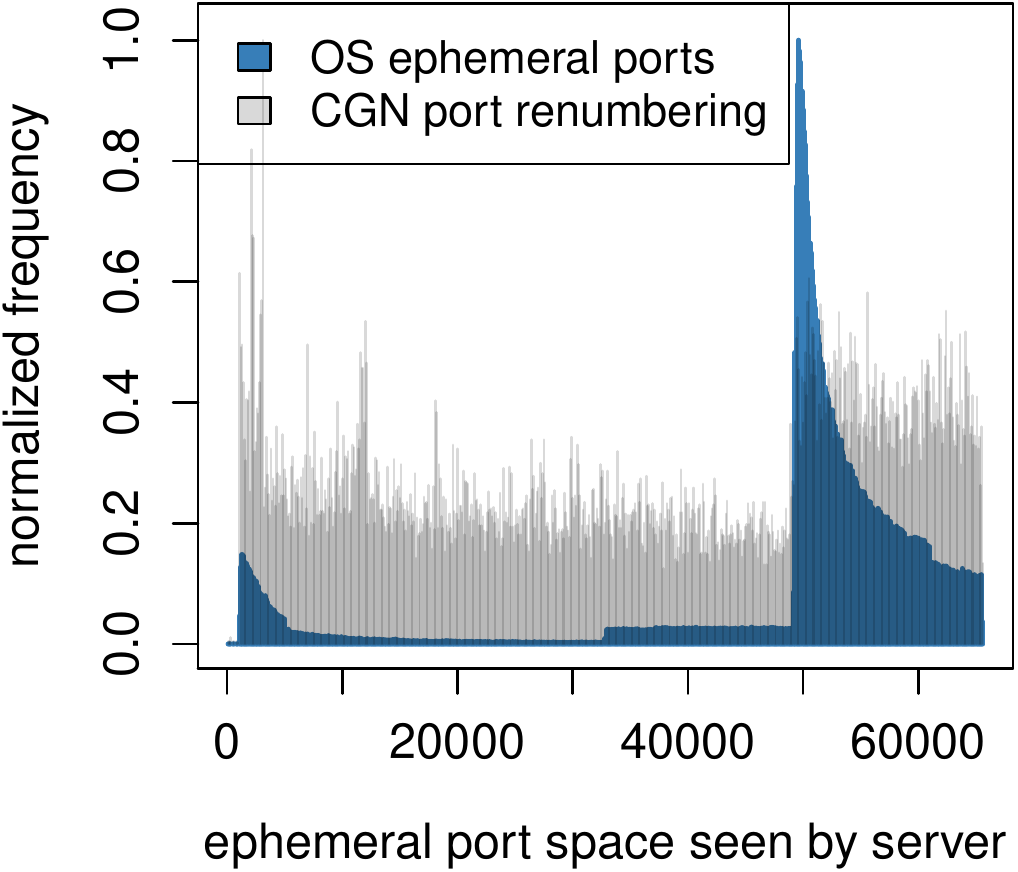}
    \label{fig:portsephemeralspace}
  }
  \hfill
  \subfigure[Port preservation behavior of CPE equipment. 92\% of UPnP sessions 
  are 
  from devices behind port-preserving CPEs.]{
    \includegraphics[width=0.3\linewidth]{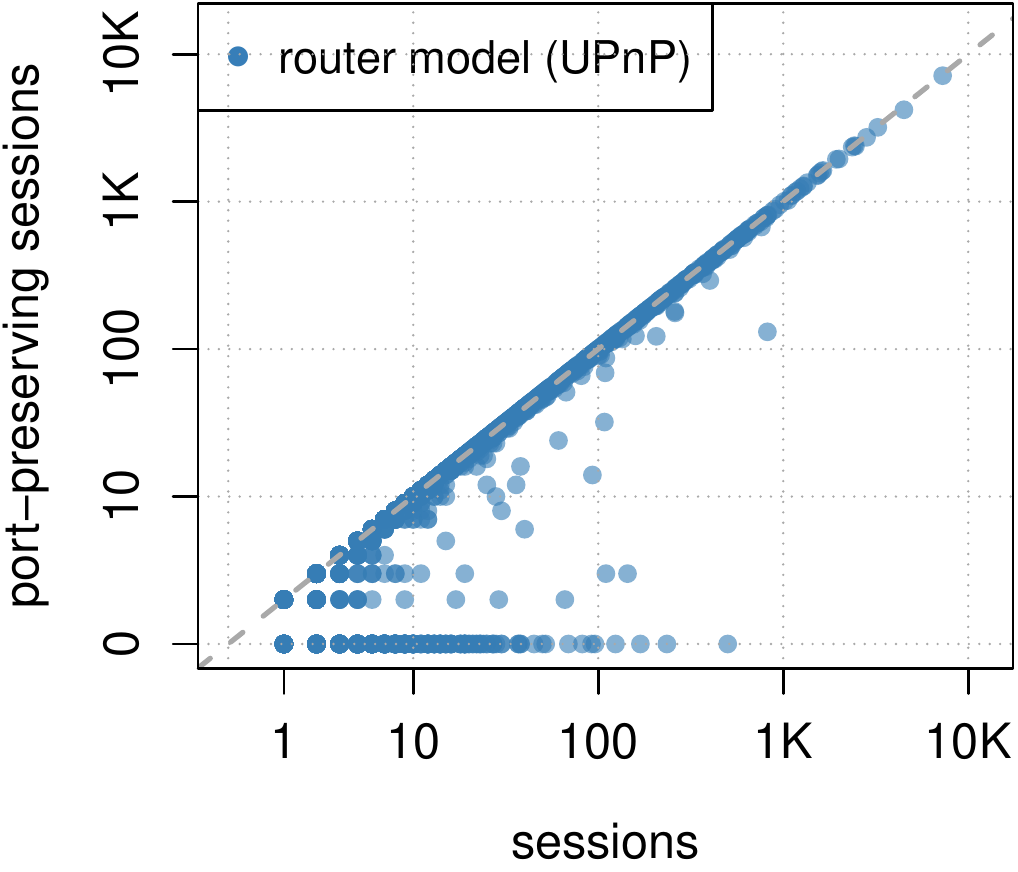}
    \label{fig:portpreservationcpe}
	
  }
  \hfill
   \subfigure[Observed ports per session, chunk-based random allocation 
   strategy 
   (4K ports per subscriber, AS12978).]{
      \centering
      \includegraphics[width=0.3\linewidth]{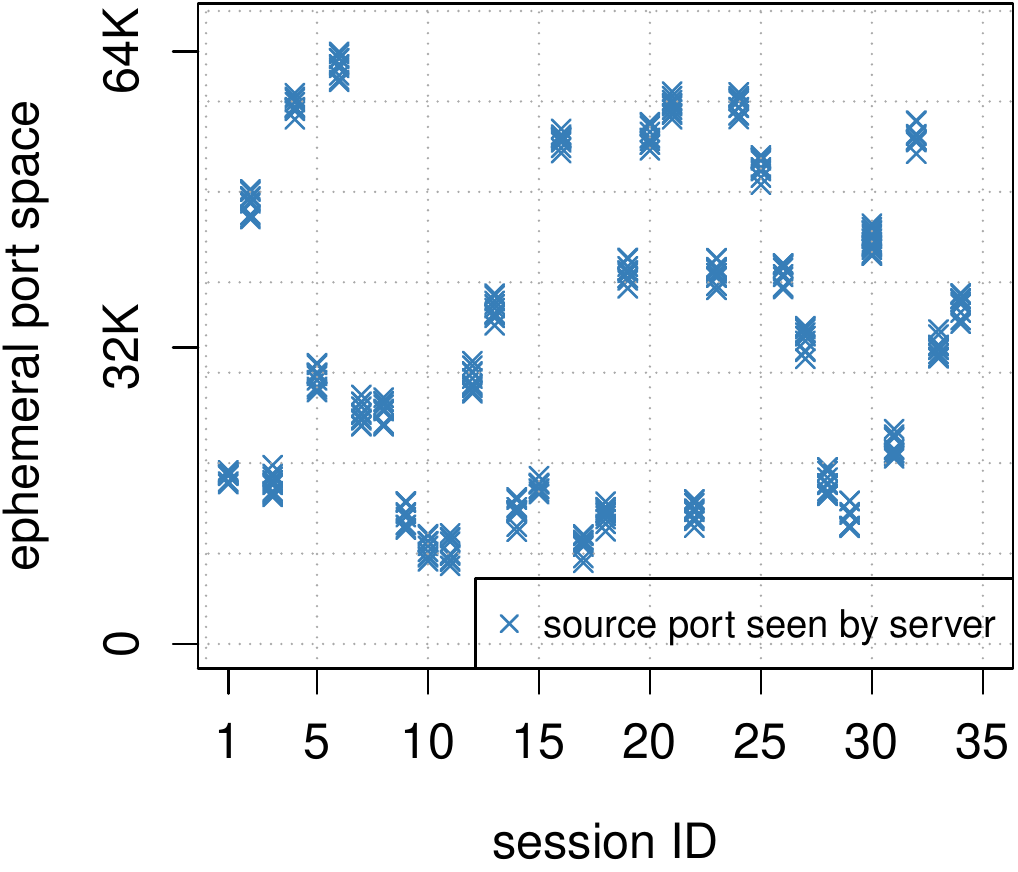}
      \label{fig:portchunks}
    }
  \caption{Port Allocation Properties.} 
  \label{fig:portproperties}
\end{figure*}

\section{Drilling into CGN Properties}
\label{sec:drilling}

Having a broad perspective of CGN deployment in today's Internet in hand, we 
next drill into the properties of the detected CGNs. NATs can be configured in 
a multitude of ways and as our survey results indicate, configuring a NAT at 
carrier-scale presents a massive resource distribution problem, including
\first public IP addresses, \second private IP addresses and 
\third ephemeral port numbers. The CGN creates state in the form of NAT 
mappings with finite lifetimes (timeouts) to associate these resources
depending on the NAT mapping type. A CGN's configuration directly
affects \first the degree of resource sharing, i.e., how many
subscribers can reside behind a given set of public IP addresses, as
well as \second the number of simultaneous flows available to
individual subscribers.

In this section, we study the configuration of our identified CGNs. In 
particular, we study \first which address ranges ISPs use internally, \second 
how CGNs assign IP addresses and ports to their subscribers, \third topological 
properties of CGNs (i.e., the location of the NAT), and \fourth the kind of NAT 
mappings deployed CGNs commonly employ. Where appropriate, we contrast
findings for CGNs with our findings for commonly deployed CPE devices.

\subsection{Internal Address Space Usage}
\label{sec:privateaddressspaces}

Our two probing methods enable us to evaluate properties of the
address space behind detected CGNs. Figure \ref{fig:cgnaddresspacebar}
shows per AS the internal address space ranges used within non-cellular as
well as cellular CGNs. Overall, we observe that naturally the 
largest private range (10X) is the most commonly used space for CGNs, followed
by the 100X block newly allocated specifically for CGN 
deployments \cite{rfc6598}. We
also observe CGNs deploying the smaller 172X and 192X address
spaces. Interestingly, roughly 20\% of the ASes use multiple ranges of
reserved address space in their CGN deployment. We speculate that the
size of individual blocks does not suffice or, more likely, that such
private address space is already in use in other parts of the
organization. Some cellular ISPs in fact use routable address space
for their internal CGN deployments. In Figure
\ref{fig:routablespaceases} we show which routable address blocks make
up the most prominent cases we detected. While most of the routable
address space used is not routed in practice (such as the 25.0.0.0/8
block, allocated to the UK Ministry of Defense), some ISPs use address
space within their internal deployment (e.g., 1.0.0.0/8) that is
publicly routed by other ASes. We contacted a representative of one of
these major ISPs who confirmed that their internal deployment of routable 
address space results from scarcity of internal address space. Thus, some ISPs 
evidently \textit{experience a shortage of internal address space} and adopt 
drastic measures at the expense of potential security and connectivity 
problems once public and internal addressing collides. Moreover, this address 
range proliferation renders troubleshooting CGNs even more cumbersome.

\subsection{Port \& IP Address Allocation}
\label{sec:ipportallocation}

Next, we study how CGNs allocate ports and IP addresses to their subscribers. 
We start with the former. NAT port allocation may adopt the following
strategies~\cite{rfc4787}: $(i)$ {\em port preservation}, where the NAT attempts to 
maintain the local port of the flows; $(ii)$ {\em sequential use}, where the NAT 
allocates ports in a sequential order; $(iii)$ {\em random use}, where the NAT 
allocates ports without a clearly identifiable pattern.\footnote{We allow some 
leeway in determining port behavior. For example, we identify port preservation
if at least 20\% of ports remain preserved, and we declare \textit{sequential 
use} if every two subsequent connections exhibit a numerical port difference 
smaller than 50. This accounts for situations in which NATs can not allocate 
the original or subsequent sequential port because of already existing 
mappings.}

\parax{Measuring port translation} During one execution of \neta (a
``session''), its client opens 10 sequential TCP flows to an
echo server listening on a high port number unlikely to be
proxied. These TCP flows enable us to reason about the port
allocation strategy implemented by the CGN, by comparing the local
ephemeral port number, as chosen by the device, with the source port
as seen by our server. Figure \ref{fig:portsephemeralspace} shows the
distribution of source port numbers as observed by our server. We show
two histograms, one for port-preserved sessions and one for sessions
exhibiting port translation. While operating systems employ ephemeral
port ranges~\cite{ephemeralportslist}, CGNs translating port numbers
utilize the entire port space. This observation could prove useful for
server-side attempts (e.g., by content providers) to identify whether
a client-side IP address belongs to a CGN.

\parax{Port translation of CPE routers} In non-cellular networks, where users' 
packets typically get NATed by a CPE, our measurements might be affected by 
CPEs employing port translation. To assess the impact of CPEs on port translation, 
Figure \ref{fig:portpreservationcpe} shows for each CPE model (inferred 
using UPnP) the number of \textit{non-CGN} sessions where our server saw the 
same ports as chosen by the device. We observe that in more than 92\% 
of non-CGN sessions the CPE did {\it not} alter the source port numbers. Hence, 
while some CPE do translate ports, their effect on our analysis remains small.

\begin{figure}
  \begin{center}
    \includegraphics[width=0.95\linewidth]{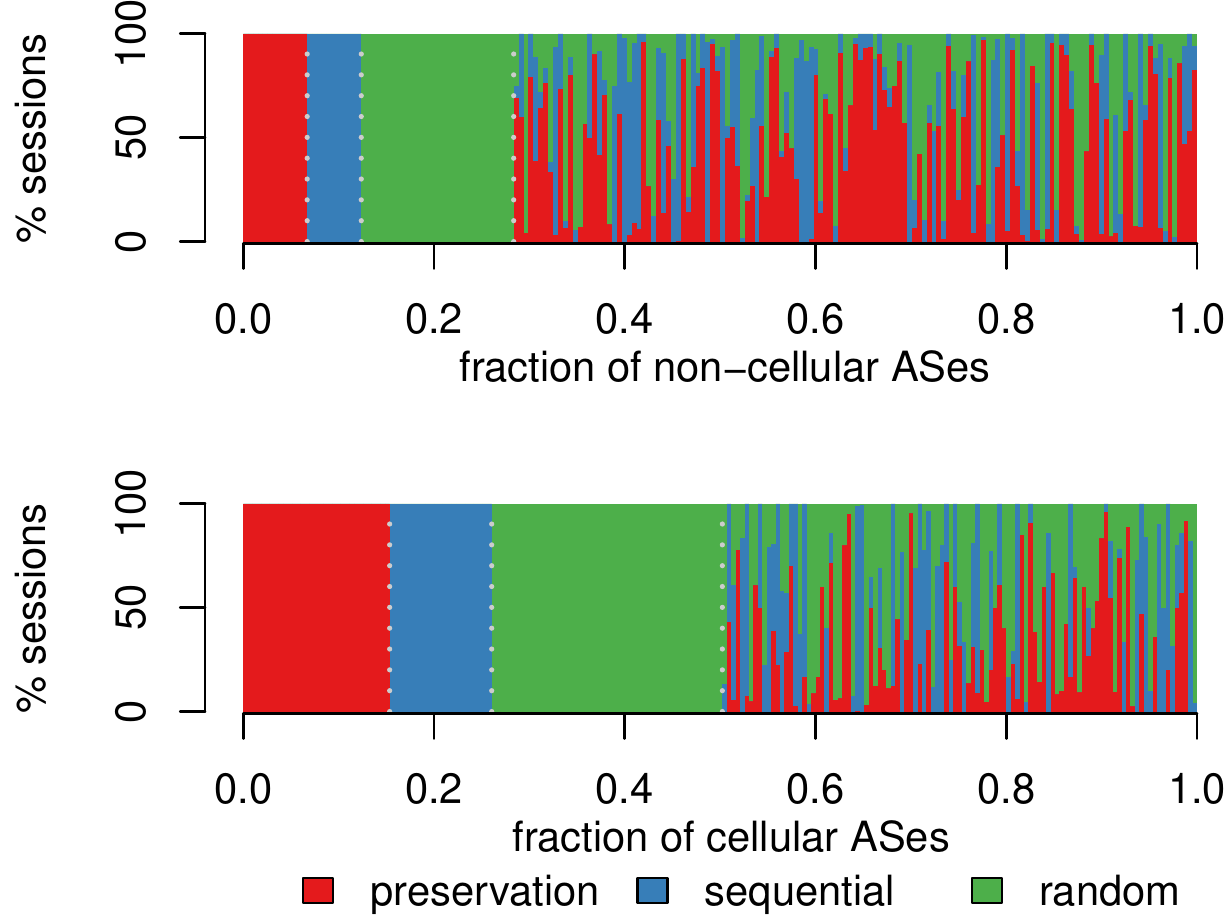}
  \caption{Distribution of observed port allocation strategies per CGN AS.}
    \label{fig:portbinginpolicies}
  \end{center}
\end{figure}

\parax{Network-wide port allocation strategies} 
Figure~\ref{fig:portbinginpolicies} shows the distribution of port allocation
strategies for each CGN-positive AS. We sort ASes with a ``pure'' 
allocation strategy (left part of the plot) and move ASes with mixed allocation 
strategies to the right. We observe a uniform port allocation strategy
for about a third of the non-cellular ASes and for about 50\% of the cellular ones.
For the rest, \textit{CGN behavior is heterogeneous.} We can attribute this to
distributed CGN deployments, where users of the same ISP reside behind
multiple CGNs, and the fact that NAT devices do not necessarily 
behave consistently, changing their behavior under load and over time 
\cite{rfc5780}. Table~\ref{table:portmapping} summarizes the dominating 
port allocation strategy per AS.

\parax{Chunk-based port allocation} In addition to the classification in our 
three categories of port allocation strategies, we also identify CGNs with 
random chunk-based allocation, where each subscriber receives a fixed port
block~\cite{cisconat}. Given sufficient data, we can infer the size of such port
``chunks''. Figure \ref{fig:portchunks} shows an example of chunk-based port 
management: AS12978 allocates 4K ports per subscriber. For each recorded session
in this AS, our server observes source ports translated in no particular order 
(neither preserved nor sequential) and that all ports of a given session fall
within a well-defined range.

In order to detect chunk-based allocation for all ASes, we require \first 
at least 20 sessions with random port translation
and \second all sessions exhibiting port numbers within a 
range smaller than 16K ports. Using this approach we identified 17 ASes 
using chunk-based allocations, shown in Table 
\ref{table:portmapping}. While a minority of the identified CGNs, it
allows us to reason directly about the dimensioning of the CGN e.g. in
terms of the number of subscribers sharing a given IP address.
We find 6 ASes in which subscribers only receive a port chunk smaller
than 1K; for 3 of them, the chunk size falls to 512 ports---a scarily
small number given that loading a single Web page can result in many
dozens of TCP connections to fetch its various objects
\cite{butkiewicz2011understanding}, resulting in a sizeable overall
number of concurrent connections in residential networks
\cite{alcock2010investigating}. The size of the port chunks then also
directly translates into the maximum number of users per public IP
address: we find 64 subscribers per IP address in the case of a 1K
port chunk.

\parax{NAT pooling behavior} For the majority of the CGN-positive ASes, we 
observe \textit{paired pooling behavior}, i.e., subsequent TCP sessions are 
bound to the same public IP address (recall \xref{sec:background}),
with varying port allocation patterns. However, we find that 21\% of the 
CGNs also employ \textit{arbitrary pooling behavior}, i.e.,  \neta reported
multiple global IP addresses during the duration of the test for more
than 60\% of the sessions. This list includes major ISPs spanning all 
geographic regions.
IETF guidelines~\cite{rfc4787} discourage this behavior due to its
complicating effect on applications (particularly SIP and RTP), which
use multiple ports on the same end host but do not negotiate IP
addresses individually~\cite{Johnston:2009:SUS:1804456, rfc4787}.

\begin{table}
\footnotesize
\resizebox{\linewidth}{!}{
\begin{tabular}{l l c c}
\multicolumn{2}{l}{\textbf{Port allocation strategy}}& \textbf{Non-cellular } & 
\textbf{Cellular} \\  \toprule
\multicolumn{2}{l}{Port-preservation} & 
41.2\%~ & 
27.9\%~ \\ 
\multicolumn{2}{l}{Sequential} & 22.2\%~ 
& 26.0\%~ \\ 
\multicolumn{2}{l}{Random}  & 35.6\%~ &  
44.7\%~ 
\\ 
\toprule
\multicolumn{2}{l}{Random (with chunk allocation)} & 
9~ 
(4.6\%{})  & 
8~ 
(3.7\%) \\  
\midrule
 \multirow{3}{*}{Chunk size (CS)} & CS $\leq$ 1K & 
 4~ & 
 2~  \\
	 & 1K $<$ CS $\leq$ 4K & 
	 2~ &  
	 3~ 
	 \\
	 & 4K $<$ CS $\leq$ 16K & 
	 3~ & 
	 3~   \\
\end{tabular}
}
\caption{Port allocation strategies observed for CGN ASes. For ASes
  implementing chunk-based random port allocation we estimate the
  per-subscriber chunk size.
}
\label{table:portmapping}
\end{table}

\subsection{CGN-specific measurements}

To extract additional details about the CGNs under study, we extended our \neta 
test suite with two tests. In this section we describe the tests'
operation; the corresponding findings follow subsequently.

\begin{figure}
  \begin{center}
    \includegraphics[width=0.99\linewidth]{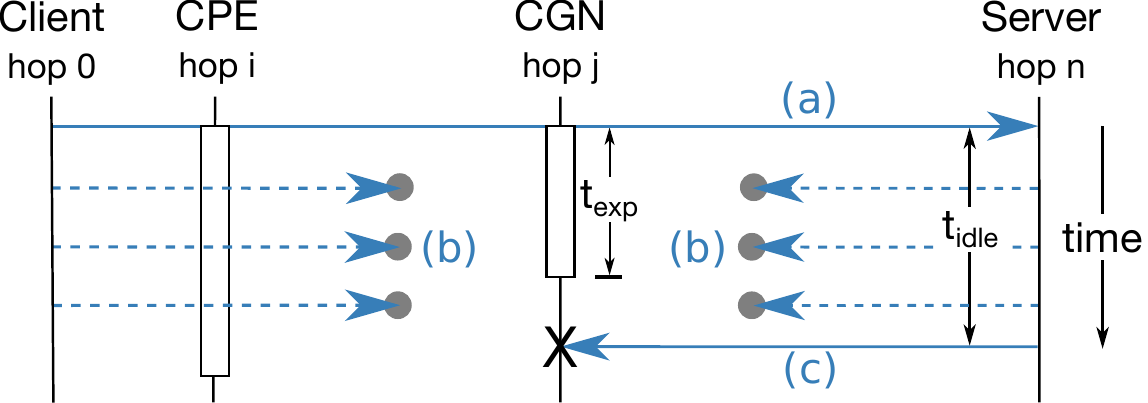}
  \caption{Reachability experiment scenario, consisting of 
  initialization packet (a), keepalive packets (b), and probe packet (c). 
  This example uses the parameters $i \leq ttl_c < j$ and $ttl_s < n-j$.
  If $t_{exp} < t_{idle}$ the mapping in the CGN (hop $j$) expires 
  before the server replies with a probe packet.}
    \label{fig:ttltestscenario}
  \end{center}
\end{figure}

\parax{TTL-driven NAT enumeration} An extension of previous
work~\cite{muller2013analysis}, this method identifies the precise
on-path location and the mapping timeouts of cascaded NATs. To do so,
it leverages the stateful nature of NATs and their need to remove the
state of idle UDP flows from the translation table (recall
\xref{sec:background}).
During the test we repeatedly perform a reachability experiment that 
selectively detects stateful middleboxes in a chosen subset of the 
path, as depicted in Figure~\ref{fig:ttltestscenario}. As annotated, the test 
consists of three stages:
\textit{(a)} the client creates a UDP flow to our server,
\textit{(b)} both endpoints transmit TTL-limited ``keepalive'' probes
for an idling period $t_{idle}$ in order to keep the flow's state
alive up to but not at the hop under test,
\textit{(c)} the server checks whether it can still reach the client.
If not, we conclude that the hop under test is a NAT that has removed
the flow's state.
Our test enumerates the 
path between client and server by systematically performing iterations of this 
reachability experiment using different parameters for the client-side initial 
TTL $ttl_c$, server-side initial TTL $ttl_s$, and elapsed idle time $t_{idle}$.

We acknowledge three limitations of this approach. First, as a
crowd-sourced test relying on user involvement, we need to limit the
idle period of the test. We test idle times up to 200 
seconds, the maximum possible value without prolonging the overall 
runtime of a \neta test session. Hence, NATs with a mapping timeout larger than 
200~seconds go unnoticed, leading to an 
underestimation of the actual number of NATs: in 
30.9\%{} of the tests (see Table~\ref{table:ttltestnumbers}) 
we do not find an expired 
mapping, while a mismatch between the client's local and server-perceived IP addresses
(\xref{sec:cgndetection:neta}) evidently indicates NAT deployment.
In the 
following, we only consider cases in which we could successfully observe an 
expired mapping.
Second, based on the results of the reachability experiment we cannot distinguish 
between NATs and other stateful middleboxes such as stateful firewalls. 
However, we find stateful middleboxes without address translation 
in only less than 0.5\%{} of our tests (see 
Table~\ref{table:ttltestnumbers}). We 
exclude these cases in the following analysis.
Third, for a reliable expiration of the keepalive packets, the technique 
requires stable path lengths. Due to the large number of reachability 
experiments per test session ($\sim$60), we can detect and filter results with 
unstable paths.

\begin{table}
\center 
\footnotesize
\resizebox{\columnwidth}{!}{
\begin{tabular}{l c c}
 & \textbf{CGN detected} & \textbf{No CGN detected} \\
\toprule
\textbf{IP address mismatch} & 67.6\%{} & 30.9\%{} \\
\textbf{IP address match} & 0.5\%{} & 0.9\%{} \\
\end{tabular}
}
\caption{Detection rate of TTL-driven NAT enumeration.}
\label{table:ttltestnumbers}
\end{table}

\parax{STUN test} To study the mapping types of CGNs, we implemented a 
STUN~\cite{rfc5389} test in our Netalyzr test suite in October 2015. 
STUN determines the mapping type implemented by on-path NATs. STUN sends probe 
packets to a public STUN server (which answers certain probe packets from a 
different port and/or IP address) and waits for the respective 
replies.\footnote{For more details on the operation of STUN, we refer 
to~\cite{rfc5389}.}

From the TTL-driven NAT enumeration test (deployed in September 2014) we have 
collected more than 38K{} sessions, whereas the STUN
test (deployed in October 2015) produced 23K{} 
sessions. 
To be able to contrast sessions from within CGN-positive networks against 
CGN-negative ones, we augment the results from both tests with the 
results from our CGN detection tests (\xref{sec:cgndetection}).
We further apply filtering rules to the results of both tests to ensure 
that we have collected at least three sessions from a particular network 
(combination of AS number and CGN classification type, e.g. 
``cellular CGN''). After applying the 
filters, this leaves us with 18K{} sessions from the NAT 
enumeration test running via both non-cellular (70\%{})
and cellular networks (30\%{}).
The results cover 608{} ASes, whereof 43\%{} 
(259{} ASes) deploy CGN.
For the STUN test we count 20K{} sessions from non-cellular
(87\%{}) and cellular networks (13\%{}).
The STUN results span 720{} ASes including 170{} 
CGN-positive ASes (24\%{}).

\subsection{Topological Properties of CGNs}
\label{sec:cgnlocation}

\begin{figure}
  \begin{center}
    \includegraphics[width=0.99\linewidth]{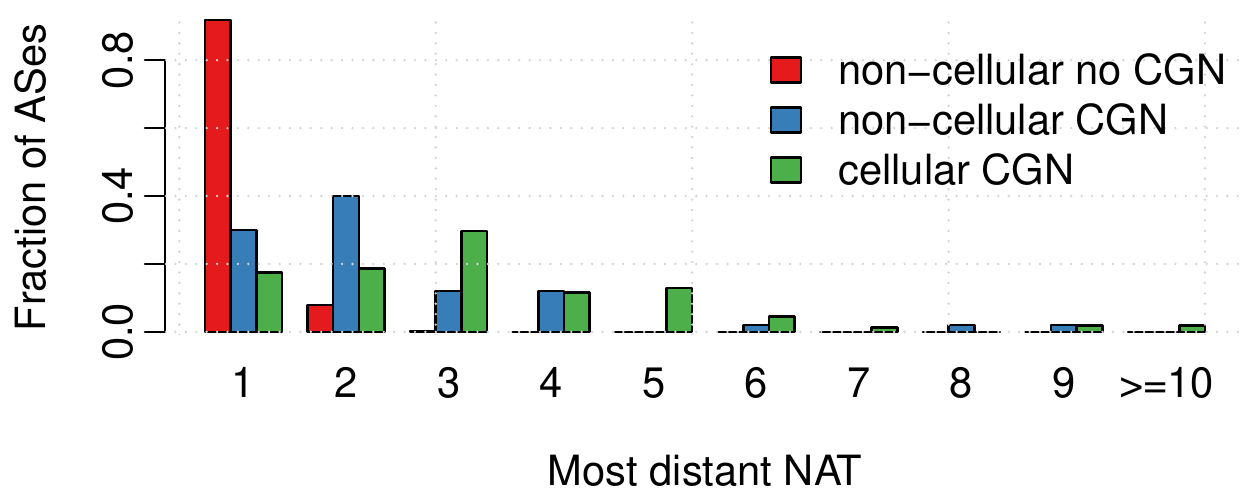}
  \caption{Maximum NAT distance from the subscriber.}
    \label{fig:natdistance}
    \vspace{-1em}
  \end{center}
\end{figure}

Figure~\ref{fig:natdistance} shows the distribution of the number of hops 
between the client and the most distant NAT detected, grouped per AS and its 
respective CGN deployment status. We detected NATs as far away as 
18{} hops from the client. As expected, most of the NATs 
in CGN-negative ASes (92\%{}) 
sit just one hop away from the client, i.e. they are typically located right on 
the CPE router. Compared to that, most CGNs are located two to five hops 
away from the client (64\%{} of 
non-cellular and 73\%{} of cellular 
ASes). In non-cellular ASes the CGN distance mostly ranges from two hops up 
to six hops. In the case of cellular ASes, however, the CGN distance ranges 
from one hop to two hops and up to 12 hops away from the client. In fact, we 
find that for 10\%{} of the cellular ASes, 
the CGN is located six or more hops away from the client. A large number of 
hops between client and CGN hints at a centralized CGN infrastructure with 
large aggregation points, which has the potential of affecting the accuracy of 
IP geolocation databases when locating the external IP address of 
clients behind CGN.

\begin{figure}
  \begin{center}
    \includegraphics[width=0.99\linewidth]{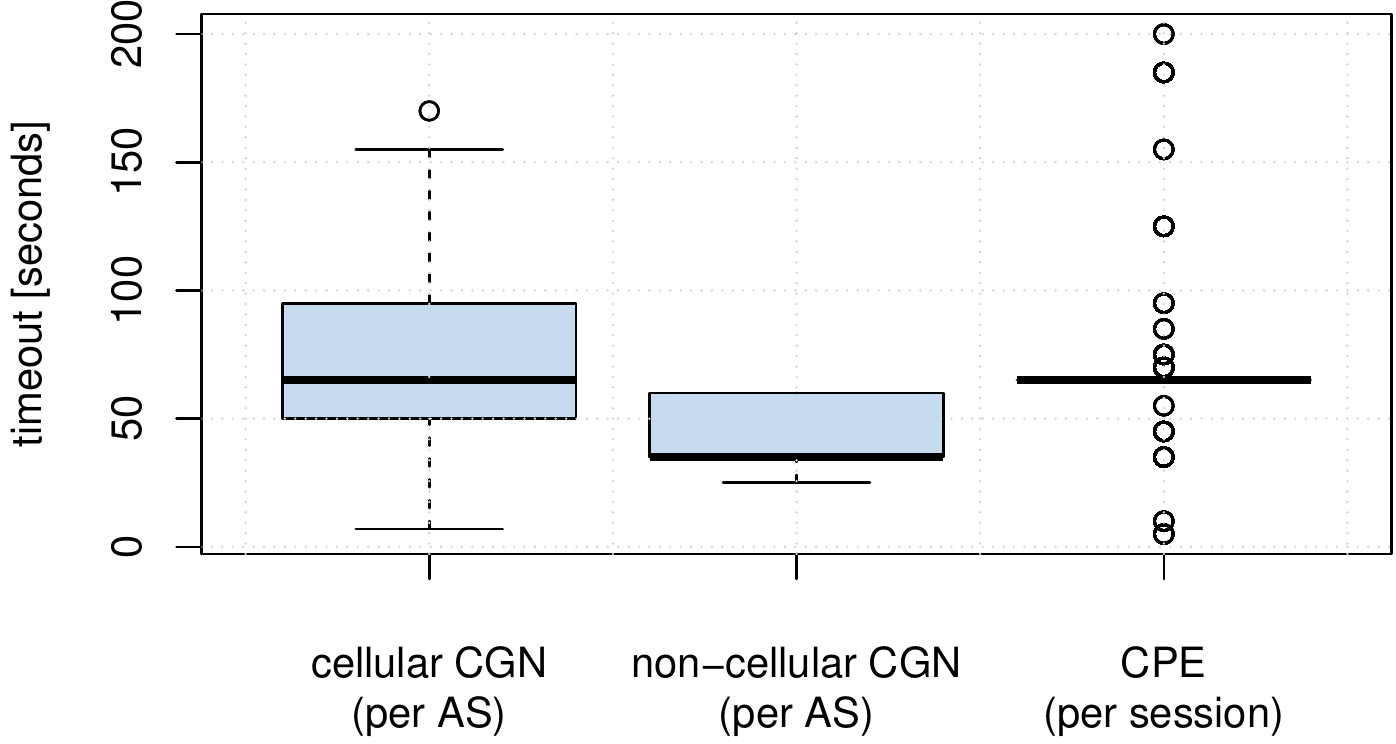}
  \caption{UDP mapping timeouts of CPEs and CGNs.}
  \vspace{-1em}
    \label{fig:nattimeout}
  \end{center}
\end{figure}

\subsection{Flow-Mapping Properties of CGNs}
\label{sec:flowmappingprops}

The type of NAT mapping (recall \xref{sec:background}) as well as its
state-keeping timeout directly affect the reachability of a host
located behind a NAT, and thus has a profound effect on applications that rely
on peer-to-peer connectivity \cite{fordHolePunching, d2009measurement} or
long-lived sparse flows~\cite{Wang:2011:USM:2043164.2018479}.

\parax{Mapping Timeouts}
Figure~\ref{fig:nattimeout} shows the UDP mapping timeouts for the 
detected CGNs, both in the cellular, as well as in the non-cellular case. Here, 
we aggregate our CGN-positive sessions on a per-AS level. An AS is 
represented by its most frequent timeout value (mode).
We also report timeout values that we detected for CPE devices (shown in the 
right boxplot), where we show a boxplot of all recorded sessions. In NAT444 
scenarios (non-cellular CGN) we need to make sure to report the timeout of CGNs 
rather than the CPE NATs. Therefore, to reason about CGN mapping timeouts, we 
only consider sessions that were detected as CGN (\xref{sec:cgndetection}) and 
where our TTL-driven NAT enumeration detected the NAT at a distance of three or 
more hops away from the client.
We observe that 74\%{} of detected NATs expire idle UDP 
state after 1 minute or less, but we find values ranging from 10s to 
200s.\footnote{Note that our timeout detection mechanism uses a 10 second 
probing interval. Hence, reported values can differ up to 10 seconds from the 
actual NAT timeout.}
CGNs in cellular networks exhibit a larger median mapping timeout 
(65{}s) compared to non-cellular networks 
(35{}s). For CPE NATs we predominantly measured 
a timeout of 65{}s. We find higher variability and 
a lower median of timeout values for non-cellular CGNs when compared to CPE 
NATs. Low CGN timeout values might in turn negatively affect the longevity of 
sparse UDP flows that are also exposed to CPE NATs. While we find lower timeout 
values for CGNs compared to CPEs, we acknowledge that this property does not 
necessarily hold true for CGNs in general, as our test can not detect timeout 
values larger than 200 seconds.

\begin{figure}
  \begin{center}
  \subfigure[Distribution of observed STUN types in CPE NATs.]{
    \includegraphics[width=0.95\linewidth]{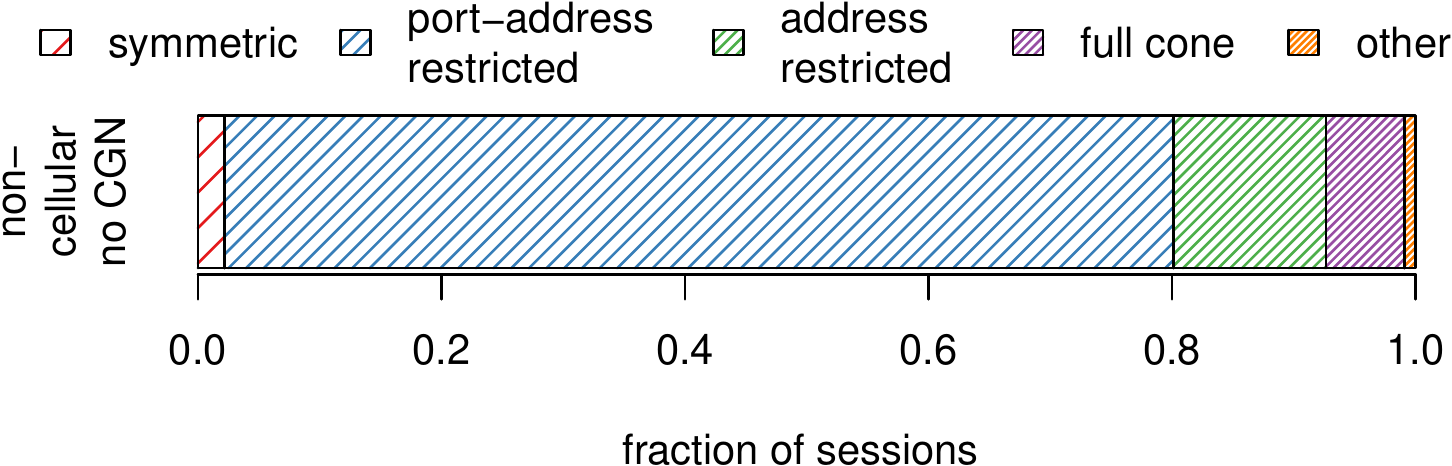}
    \label{fig:stunnocgn}
  }
  \subfigure[Most permissive STUN type per AS (only CGN sessions).]{
    \includegraphics[width=0.95\linewidth]{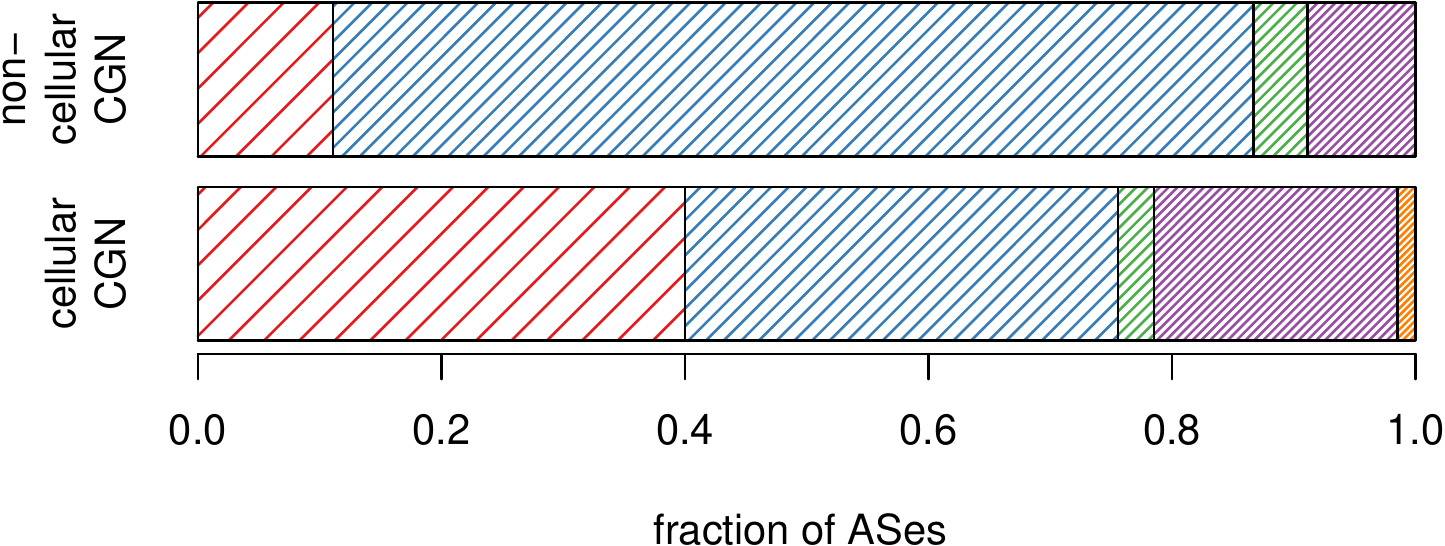}
    \label{fig:stuntypes}
  }
  \caption{STUN results per AS.}
  \vspace{-1em}
  \label{fig:stun}
  \end{center}
\end{figure}

\parax{Mapping types}
Figure~\ref{fig:stun} shows our STUN results. We order the observed 
mapping types from most restrictive (\textit{symmetric NAT}) to most permissive 
(\textit{full cone NAT}).
In Figure \ref{fig:stunnocgn} we show the NAT mapping type as observed
for CPE routers, while the bars in Figure \ref{fig:stuntypes} indicate
the most permissive type of NAT mapping for our CGN-positive ASes.
Recall that when multiple NAT devices reside on the path, STUN reports the most restrictive behavior of them, which also determines 
eventual NAT traversal. Hence, we argue that the most permissive STUN type 
provides a good approximation for the CGN behavior, 
because there cannot be a STUN result less restrictive than the CGN. 
We observe that, while exhibiting some diversity, less than 2\% 
of the tests showed CPE NATs with very 
restrictive symmetric NATs. In contrast to CPE NATs, we observe 
11\%{} of non-cellular CGN ASes whose most permissive 
mapping type is symmetric. Among these networks we find many 
popular large European ISPs.
For cellular networks we observe a bimodal outcome, with a 
large fraction of both restrictive (40\%{} 
symmetric) and permissive (20\%{} full cone) NAT 
types. We see large operators on both ends of the spectrum, with major cellular 
networks in the US deploying CGNs with symmetric mapping types.

Thus, we often measure stricter NAT mapping policies for CGN-positive sessions 
when compared to common home CPE devices. We conclude that a large fraction of 
ISPs deploy CGNs that use symmetric flow mappings, which limits the customers' 
ability to establish direct connections. For this reason, the IETF lists an 
endpoint-independent mapping (which symmetric NATs violate) as their first 
requirement for CGNs~\cite{rfc4787,rfc5382}.

\section{Implications} %

Our analysis shows that ISPs widely deploy CGN. We find that more than 17\% 
of eyeball ASes and more than 90\% of cellular ASes rely on CGNs 
(\xref{sec:networkwideview}), with particularly high deployment rates in Asia 
and Europe---regions in which IPv4 address scarcity cropped up first, as the 
respective registries ran out of readily available IPv4 addresses in 2011 and 
2012. Thus, adopting CGN presents
a viable alternative to buying IPv4 address space from brokers.
CGNs actively extend the lifetime of IPv4 and hence also fuel the 
demand of the growing market for IPv4 address space \cite{RABP14}, which in 
turn affects market prices and possibly hampers the adoption IPv6.

CGNs directly affect ``how much Internet'' a subscriber receives, by \first
limiting available ephemeral port space, \second restricting 
the directionality of connections, and \third limiting connection 
lifetimes due to finite state-keeping budgets.  Studying our identified CGN 
deployments, we find 
a wide spectrum of configurations and degrees of address sharing 
(\xref{sec:drilling}). On the limiting end of the spectrum, we find ISPs 
allocating as little as 512 ephemeral ports per subscriber
(\xref{sec:ipportallocation}), multiplexing up to 128 subscribers per public 
IP address. Comparing NAT flow mapping types and timeouts of CGNs to commonly 
deployed CPE hardware, we find that in many instances CGNs use more 
restrictive flow mapping types when compared to their home counterparts 
(\xref{sec:flowmappingprops}). This rules out peer-to-peer connectivity,
complicating modern protocols such as WebRTC~\cite{rfc7478} that now need to
rely on rendezvous servers.

We argue that the lack of \textit{guidelines} and \textit{regulations}
for CGN deployment compounds the situation. While the IETF publishes
best practices for general NAT behavior \cite{rfc4787,rfc5382,rfc7857}
as well as basic requirements for CGN deployments \cite{rfc6888}
(which, incidentally, many of our identified CGNs violate),
dimensioning NATs at carrier-scale in a way that minimizes collateral
damage remains a black art. Our finding that some large ISPs find the
need to employ publicly routable (indeed, sometimes routed) address
space for internal CGN deployment (\xref{sec:privateaddressspaces})
underlines the graveness of the situation.
While it remains out of scope for this work to precisely measure the
effect of CGNs on end-users' applications, we believe that our
observations can serve as input for establishing such guidelines.
Our findings should also interest regulators, who in some countries
already impose acceptable service requirements on Internet performance
(e.g., the FCC's measurements of advertised vs. achieved throughput
\cite{johnston2013measuring}).  We argue that the presence and service
levels of CGNs should be readily identifiable in ISPs'
offerings. Unfortunately, we find that most ISPs do not cover CGN
deployment in their terms of service.
Lastly, our findings document further erosion of the meaningfulness of
IP address reputation, address-based blacklisting, IP-to-user
attribution, and geolocating end-users (\xref{sec:ipportallocation},
\xref {sec:cgnlocation}), which become all but infeasible in the
presence of CGN.

\section{Conclusion}
This work presents a solid first step towards a better understanding
of the prevalence and characteristics of CGN deployment in today's
Internet. Our methods, based on harvesting the BitTorrent DHT and
extensions to our \neta active measurement framework, prove effective
in uncovering CGN deployments: we detect more than 500 instances in
ISPs around the world. When assessing the properties of these CGNs we
find striking variability in the dimensioning, configuration,
placement, and effect on the subscriber's connectivity. We hope this
study will stimulate a much-needed discussion about best practices,
guidelines, and regulation of CGN deployment. In future work, we plan
to further improve our detection mechanisms and to study the impact of
CGN on application performance.

\section*{Acknowledgments}
\label{sec:acks}

We thank the network operators who participated in our survey for their 
insightful feedback and comments. We thank Daniele Iamartino for support with 
validating the BitTorrent data, and Martin Ott for his support with implementing
the BitTorrent crawler. We also thank the anonymous reviewers for their helpful 
feedback. This work was partially supported by the
US National Science Foundation under grants CNS-1111672 and CNS-1213157,
the Leibniz Prize 
project funds of DFG/German Research Foundation (FKZ FE 570/4-1), and
the BMBF AutoMon project (16KIS0411).  Any
opinions expressed are those of the authors and not necessarily of
the sponsors.

\bibliographystyle{plain}
\bibliography{paper}

\end{document}

%% file: preamble.tex
\def\first{({\it i})\xspace}
\def\second{({\it ii})\xspace}
\def\third{({\it iii})\xspace}
\def\fourth{({\it iv})\xspace}

\def\TblSpB{\rule{0pt}{3ex}}

\providecommand{\ie}{{i.e.}, }
\providecommand{\eg}{{e.g.}, }

\providecommand{\etal}{\emph{et al.}\xspace}

\newcommand{\privateipport}{$\text{\it IP}_{int}$$:$$\text{\it 
port}_{int}$\xspace}
\newcommand{\publicipport}{$\text{\it IP}_{ext}$$:$$\text{\it 
port}_{ext}$\xspace}
\newcommand{\dstipport}{$\text{\it IP}_{dst}$$:$${\it port}_{dst}$\xspace}

\newcommand{\privateport}{$\text{\it port}_{int}$\xspace}
\newcommand{\publicport}{$\text{\it port}_{ext}$\xspace}

\newcommand{\privateip}{$\text{\it IP}_{int}$\xspace}
\newcommand{\publicip}{$\text{\it IP}_{ext}$\xspace}
\newcommand{\dstip}{$\text{\it IP}_{dst}$\xspace}

\newcommand{\btping}{\textit{bt\_ping}\xspace}
\newcommand{\btfindnodes}{\textit{find\_nodes}\xspace}

\newcommand{\btnodeid}{\textit{nodeid}\xspace}

%% file: paper.bbl
\begin{thebibliography}{10}

\bibitem{netalyzrGPlay}
{Netalyzr for Android. Google Play}.
\newblock
  \url{https://play.google.com/store/apps/details?id=edu.berkeley.icsi.netalyzr.android}.

\bibitem{cgn_menog}
{A.~Tabdili}.
\newblock {Carrier Grade NAT: Requirements and Challenges in the Real World }.
\newblock
  \url{http://www.menog.org/presentations/menog-10/Amir%20Tabdili%20-%20Carrier%20Grade%20NAT.pdf}.

\bibitem{a10nat}
{A10 Networks}.
\newblock {Carrier Grade NAT (CGN) Deployment Guide}.
\newblock
  \url{https://www.a10networks.com/sites/default/files/resource-files/A10-DG-Carrier_Grade_NAT_%28CGN%29_Large_Scale_NAT_%28LSN%29.pdf}.

\bibitem{alcock2010investigating}
S.~Alcock, R.~Nelson, and D.~Miles.
\newblock {Investigating the Impact of Service Provider NAT on Residential
  Broadband Users}.
\newblock {\em TR, University of Waikato}, 2010.

\bibitem{aspop_apnic}
{APNIC Labs}.
\newblock {Customers per AS Measurements}.
\newblock Description: \url{https://labs.apnic.net/?p=526} Dataset:
  \url{http://stats.labs.apnic.net/aspop}.

\bibitem{rfc4787}
F.~Audet and C.~Jennings.
\newblock {Network Address Translation (NAT) Behavioral Requirements for
  Unicast UDP}.
\newblock RFC 4787 (Best Current Practice), January 2007.
\newblock Updated by RFCs 6888, 7857.

\bibitem{nat_detection_ipid:imw}
S.~M. Bellovin.
\newblock {A Technique for Counting NATted Hosts}.
\newblock In {\em {IMW}}, 2002.

\bibitem{bittorrent_dht}
{BitTorrent.org}.
\newblock {DHT Protocol (BEP-05)}.
\newblock \url{http://www.bittorrent.org/beps/bep_0005.html}.

\bibitem{cgnimpactwebbrowsing}
E.~Bocchi, A.~S. Khatouni, S.~Traverso, A.~Finamore, V.~D. Gennaro, M.~Mellia,
  M.~Munafo, and D.~Rossi.
\newblock {Impact of Carrier-Grade NAT on Web Browsing}.
\newblock In {\em {IWCMC}}, 2015.

\bibitem{rfc6970}
M.~Boucadair, R.~Penno, and D.~Wing.
\newblock {Universal Plug and Play (UPnP) Internet Gateway Device - Port
  Control Protocol Interworking Function (IGD-PCP IWF)}.
\newblock RFC 6970 (Proposed Standard), July 2013.

\bibitem{butkiewicz2011understanding}
M.~Butkiewicz, H.~V. Madhyastha, and V.~Sekar.
\newblock {Understanding Website Complexity: Measurements, Metrics, and
  Implications}.
\newblock In {\em {IMC}}, 2011.

\bibitem{cisconat}
{Cisco}.
\newblock {NAT Administration Guide, StarOS Release 17}.
\newblock
  \url{http://www.cisco.com/c/dam/en/us/td/docs/wireless/asr_5000/17-0/PDF/17-NAT-Admin.pdf}.

\bibitem{ephemeralportslist}
Cymru.
\newblock {{Ephemeral Source Port Selection Strategies}}.
\newblock \url{https://www.cymru.com/jtk/misc/ephemeralports.html}.

\bibitem{ipv6sigcomm14}
J.~Czyz, M.~Allman, J.~Zhang, S.~Iekel-Johnson, E.~Osterweil, and M.~Bailey.
\newblock {Measuring IPv6 Adoption}.
\newblock In {\em ACM SIGCOMM}, 2014.

\bibitem{d2009measurement}
L.~D'Acunto, J.A. Pouwelse, and H.J. Sips.
\newblock {{A measurement of NAT \& Firewall Characteristics in Peer to Peer
  Systems}}.
\newblock In {\em {ASCI}}, 2009.

\bibitem{dicioccio2012probe}
L.~DiCioccio, R.~Teixeira, M.~May, and C.~Kreibich.
\newblock {Probe and Pray: Using UPnP for Home Network Measurements}.
\newblock In {\em {PAM}}, 2012.

\bibitem{rfc7021}
C.~Donley, L.~Howard, V.~Kuarsingh, J.~Berg, and J.~Doshi.
\newblock {Assessing the Impact of Carrier-Grade NAT on Network Applications}.
\newblock RFC 7021 (Informational), September 2013.

\bibitem{rfc1631}
K.~Egevang and P.~Francis.
\newblock {The IP Network Address Translator (NAT)}.
\newblock RFC 1631 (Informational), May 1994.
\newblock Obsoleted by RFC 3022.

\bibitem{johnston2013measuring}
FCC.
\newblock {Measuring Broadband America}.
\newblock \url{https://www.measuringbroadbandamerica.com/}.

\bibitem{fordHolePunching}
B.~Ford, P.~Srisuresh, and D.~Kegel.
\newblock {Peer-to-Peer Communication Across Network Address Translators}.
\newblock In {\em USENIX ATC}, 2005.

\bibitem{rfc5382}
S.~Guha, K.~Biswas, B.~Ford, S.~Sivakumar, and P.~Srisuresh.
\newblock {NAT Behavioral Requirements for TCP}.
\newblock RFC 5382 (Best Current Practice), October 2008.
\newblock Updated by RFC 7857.

\bibitem{rfc7478}
C.~Holmberg, S.~Hakansson, and G.~Eriksson.
\newblock {Web Real-Time Communication Use Cases and Requirements}.
\newblock RFC 7478 (Informational), March 2015.

\bibitem{Johnston:2009:SUS:1804456}
Alan~B. Johnston.
\newblock {\em SIP: Understanding the Session Initiation Protocol}.
\newblock Artech House, Inc., Norwood, MA, USA, 3rd edition, 2009.

\bibitem{netalyzr:imc:short}
C.~Kreibich, N.~Weaver, B.~Nechaev, and V.~Paxson.
\newblock {Netalyzr: Illuminating The Edge Network}.
\newblock In {\em {IMC}}, 2010.

\bibitem{natdection_passive:conext}
V.~Krmicek, J.~Vykopal, and R.~Krejci.
\newblock {NetFlow Based System for NAT Detection}.
\newblock In {\em {ACM CoNEXT}}, 2009.

\bibitem{nat444pam}
A.~Lutu, M.~Bagnulo, A.~Dhamdhere, and k.~claffy.
\newblock {NAT Revelio: Detecting NAT444 in the ISP}.
\newblock In {\em {PAM}}, 2016.

\bibitem{rfc5780}
D.~MacDonald and B.~Lowekamp.
\newblock {NAT Behavior Discovery Using Session Traversal Utilities for NAT
  (STUN)}.
\newblock RFC 5780 (Experimental), May 2010.

\bibitem{residential_nat:pam}
G.~Maier, F.~Schneider, and A.~Feldmann.
\newblock {NAT Usage in Residential Broadband Networks}.
\newblock In {\em {PAM}}, 2011.

\bibitem{maymounkov2002kademlia}
P.~Maymounkov and D.~Mazieres.
\newblock {Kademlia: A Peer-to-Peer Information System Based on the XOR
  Metric}.
\newblock In {\em Peer-to-Peer Systems}. Springer, 2002.

\bibitem{muller2013analysis}
A.~M{\"u}ller, F.~Wohlfart, and G.~Carle.
\newblock {Analysis and Topology-based Traversal of Cascaded Large Scale NATs}.
\newblock In {\em {ACM HotMiddlebox}}, 2013.

\bibitem{nowispnz}
{NOW (New Zealand ISP)}.
\newblock {What if I need a public IP Address?}
\newblock
  \url{https://support.nownz.co.nz/support/solutions/articles/5000504832-what-if-i-need-a-public-ip-address-}.

\bibitem{ohara2014cgnmobile}
Y.~Ohara, K.~Nishizuka, K.~Chinen, K.~Akashi, M.~Kohrin, E.~Muramoto, and
  S.~Miyakawa.
\newblock {On the Impact of Mobile Network Delays on Connection Establishment
  Performance of a Carrier Grade NAT Device}.
\newblock In {\em ACM AINTEC}, 2014.

\bibitem{rfc7857}
R.~Penno, S.~Perreault, M.~Boucadair, S.~Sivakumar, and K.~Naito.
\newblock {Updates to Network Address Translation (NAT) Behavioral
  Requirements}.
\newblock RFC 7857 (Best Current Practice), April 2016.

\bibitem{rfc6888}
S.~Perreault, I.~Yamagata, S.~Miyakawa, A.~Nakagawa, and H.~Ashida.
\newblock {Common Requirements for Carrier-Grade NATs (CGNs)}.
\newblock RFC 6888 (Best Current Practice), April 2013.

\bibitem{rfc1918}
Y.~Rekhter, B.~Moskowitz, D.~Karrenberg, G.~J. de~Groot, and E.~Lear.
\newblock {Address Allocation for Private Internets}.
\newblock RFC 1918 (Best Current Practice), February 1996.
\newblock Updated by RFC 6761.

\bibitem{RABP14}
P.~Richter, M.~Allman, R.~Bush, and V.~Paxson.
\newblock {A Primer on IPv4 Scarcity}.
\newblock {\em ACM CCR}, 45(2), 2015.

\bibitem{IMC2016Beyond}
P.~Richter, G.~Smaragdakis, D.~Plonka, and A.~Berger.
\newblock {Beyond Counting: New Perspectives on the Active IPv4 Address Space}.
\newblock In {\em ACM IMC}, 2016.

\bibitem{rfc6544}
J.~Rosenberg, A.~Keranen, B.~B. Lowekamp, and A.~B. Roach.
\newblock {TCP Candidates with Interactive Connectivity Establishment (ICE)}.
\newblock RFC 6544 (Proposed Standard), March 2012.

\bibitem{rfc5389}
J.~Rosenberg, R.~Mahy, P.~Matthews, and D.~Wing.
\newblock {Session Traversal Utilities for NAT (STUN)}.
\newblock RFC 5389 (Proposed Standard), October 2008.
\newblock Updated by RFC 7350.

\bibitem{rfc3489}
J.~Rosenberg, J.~Weinberger, C.~Huitema, and R.~Mahy.
\newblock {STUN - Simple Traversal of User Datagram Protocol (UDP) Through
  Network Address Translators (NATs)}.
\newblock RFC 3489 (Proposed Standard), March 2003.
\newblock Obsoleted by RFC 5389.

\bibitem{address_sharing_ton}
N.~Skoberne, O.~Maennel, I.~Phillips, R.~Bush, J.~Zorz, and M.~Ciglaric.
\newblock {IPv4 Address Sharing Mechanism Classification and Tradeoff
  Analysis}.
\newblock {\em IEEE/ACM ToN}, 2014.

\bibitem{pbl}
{Spamhaus}.
\newblock {The Policy Block List}.
\newblock \url{https://www.spamhaus.org/pbl/}.

\bibitem{rfc5128}
P.~Srisuresh, B.~Ford, and D.~Kegel.
\newblock {{State of Peer-to-Peer (P2P) Communication across Network Address
  Translators (NATs)}}.
\newblock RFC 5128 (Informational), March 2008.

\bibitem{wang2012real}
L.~Wang and J.~Kangasharju.
\newblock {Real-world sybil attacks in BitTorrent mainline DHT}.
\newblock In {\em IEEE GLOBECOM}, 2012.

\bibitem{Wang:2011:USM:2043164.2018479}
Z.~Wang, Z.~Qian, Q.~Xu, Z.~M. Mao, and M.~Zhang.
\newblock {An Untold Story of Middleboxes in Cellular Networks}.
\newblock In {\em {ACM SIGCOMM}}, 2011.

\bibitem{rfc6598}
J.~Weil, V.~Kuarsingh, C.~Donley, C.~Liljenstolpe, and M.~Azinger.
\newblock {IANA-Reserved IPv4 Prefix for Shared Address Space}.
\newblock RFC 6598 (Best Current Practice), April 2012.

\bibitem{nat44and444}
D.~Wing.
\newblock {NAT Tutorial}.
\newblock In {\em IETF 78}, 2010.

\end{thebibliography}
